\documentclass{statsoc}

\usepackage[colorlinks,citecolor=blue,linkcolor=blue,urlcolor=blue]{hyperref}
\usepackage{latexsym, epsfig, amssymb, amsmath, graphicx, mathrsfs}
\usepackage[left=1in,right=1in,top=1in,bottom=1in]{geometry}
\usepackage[OT1]{fontenc}

\newtheorem{lemma}{Lemma}
\newtheorem{proposition}{Proposition}
\newtheorem{theorem}{Theorem}

\newtheorem{condition}{Condition}

\newcommand{\bff}{\mbox{\bf f}}

\newcommand{\bu}{\mbox{\bf u}}
\newcommand{\bv}{\mbox{\bf v}}
\newcommand{\bw}{\mbox{\bf w}}
\newcommand{\bx}{\mbox{\bf x}}
\newcommand{\by}{\mbox{\bf y}}
\newcommand{\bz}{\mbox{\bf z}}
\newcommand{\bA}{\mbox{\bf A}}
\newcommand{\bB}{\mbox{\bf B}}
\newcommand{\bC}{\mbox{\bf C}}

\newcommand{\bF}{\mbox{\bf F}}

\newcommand{\bP}{\mbox{\bf P}}

\newcommand{\bI}{\mbox{\bf I}}

\newcommand{\bO}{\mbox{\bf O}}

\newcommand{\bS}{\mbox{\bf S}}
\newcommand{\bU}{\mbox{\bf U}}

\newcommand{\bW}{\mbox{\bf W}}
\newcommand{\bX}{\mbox{\bf X}}

\newcommand{\bZ}{\mbox{\bf Z}}
\newcommand{\bone}{\mbox{\bf 1}}
\newcommand{\bzero}{\mbox{\bf 0}}
\newcommand{\bveps}{\mbox{\boldmath $\varepsilon$}}

\newcommand{\bbeta}{\mbox{\boldmath $\beta$}}
\newcommand{\bdelta}{\mbox{\boldmath $\delta$}}

\newcommand{\bgamma}{\mbox{\boldmath $\gamma$}}

\newcommand{\bleta}{\mbox{\boldmath $\eta$}}

\newcommand{\bLambda}{\mbox{\boldmath $\Lambda$}}
\newcommand{\bSig}{\mbox{\boldmath $\Sigma$}}

\newcommand{\hbbeta}{\widehat\bbeta}
\newcommand{\hbgamma}{\widehat\bgamma}

\newcommand{\Sig}{\mathbf{\Sigma}}

\newcommand{\tr}{\mathrm{tr}}

\newcommand{\supp}{\mathrm{supp}}

\def\t{^T}

\title[NSL]
{Nonsparse learning with latent variables
\thanks{This work was supported by a grant from the Simons Foundation,  National Natural Science Foundation of China Grants 11601501, 11671018, 71532001, and 71731010, and Anhui Provincial Natural Science Foundation Grant 1708085QA02.}
}

\author{Zemin Zheng,}
\address{University of Science and Technology of China, China}
\email{zhengzm@ustc.edu.cn}
\author{Jinchi Lv}
\address{University of Southern California, Los Angeles, USA}
\email{jinchilv@marshall.usc.edu}
\author[Z. Zheng, J. Lv and W. Lin]{and Wei Lin}
\address{Peking University, China}
\email{weilin@math.pku.edu.cn}

\begin{document}

\begin{abstract}
As a popular tool for producing meaningful and interpretable models, large-scale sparse learning works efficiently when the underlying structures are indeed or close to sparse. However, naively applying the existing regularization methods can result in misleading outcomes due to model misspecification. In particular, the direct sparsity assumption on coefficient vectors has been questioned in real applications. Therefore, we consider nonsparse learning with the conditional sparsity structure that the coefficient vector becomes sparse after taking out the impacts of certain unobservable latent variables. A new methodology of nonsparse learning with latent variables (NSL) is proposed to simultaneously recover the significant observable predictors and latent factors as well as their effects. We explore a common latent family incorporating population principal components and derive the convergence rates of both sample principal components and their score vectors that hold for a wide class of distributions. With the properly estimated latent variables, properties including model selection consistency and oracle inequalities under various prediction and estimation losses are established for the proposed methodology. Our new methodology and results are evidenced by simulation and real data examples.


\end{abstract}

\section{Introduction} \label{Sec1}
Advances of information technologies have made high-dimensional data increasingly frequent particularly in the domains of genomic and metagenomic data in biology, imaging data in machine learning, and high frequency transaction data in economics. The key assumption that enables high-dimensional statistical inference is that the regression function lies in a low-dimensional manifold \citep{Hastie2009, FanLv2010, Buhlmann2011}, meaning that the model parameter vector is sparse with many zero components. Based on this sparsity assumption, a long list of regularization methods have been developed to generate meaningful and interpretable models, including \cite{Tibshirani1996, Fan2001, Zou2005, Candes2007, Sun2012, Chen2016}, among many others. Theoretical guarantees such as oracle properties, oracle inequalities, model selection consistency, asymptotic distributions, and false discovery rate control were also established for various regularization methods. See, for example, \cite{Zhao2006, Bickel2009, Tang2010, Fan12, Fan2012, Java14, Geer14, Zhang2014, Barber15, Candes16, Lee16}.

Although large-scale sparse learning works efficiently when the underlying structures are indeed or close to sparse, naively applying the existing regularization methods can result in misleading outcomes due to model misspecification \citep{White1982, Lv2014}. In particular, it was imposed in most high-dimensional inference methods that the coefficient vectors are sparse, which has been questioned in real applications. For instance, \cite{Boyle2017} suggested the omnigenic model that the genes associated with complex traits tend to be spread across most of the genome. Similarly, it was conjectured earlier in \cite{Pritchard2001} that instead of being sparse, the causal variants responsible for a trait can be distributed. Under such cases, making correct statistical inference is an important yet challenging task. Though it is generally impossible to accurately estimate large numbers of nonzero parameters with relatively low sample size, nonsparse learning may be achieved by considering a natural extension of the sparse scenario, that is, the conditional sparsity structure. Specifically, we assume the coefficient vector to be sparse after taking out the impacts of certain unobservable latent variables. A similar idea was exploited in \cite{Fan2013} by the low rank plus sparse representation for large covariance estimation, where a sparse error covariance structure is imposed after extracting common but unobservable factors.

To characterize the impacts of latent variables, various methods have been proposed under different model settings. For instance, the latent and observed variables were assumed to be jointly Gaussian in \cite{Chandrasekaran2012} for graphical model selection. To control for confounding in genetical genomics studies, \cite{Lin2015} used genetic variants as instrumental variables. \cite{Pan2015} characterized latent variables by confirmatory factor analysis (CFA) in survival analysis and estimated them using the EM algorithm. Despite the growing literature, relatively few studies deal with latent variables in high dimensions. In this paper, we focus on high-dimensional linear regression and allow for both the numbers of observable predictors and potential latent variables to be large, where the latent variables are nonsparse linear combinations of a group of observable covariates (not necessarily the predictors). To the best of our knowledge, this is a new contribution to the case of high-dimensional latent variables. Under this setup, our goals are to (i) effectively estimate latent factors given the observable covariates, (ii) identify the significant features and estimate their effects from a large pool of predictors in the presence of the underlying factors, (iii) recover the whole structure by further finding out how many important confounding factors exist as well as their impacts.


We would like to provide some partial answers to the aforementioned questions by considering an important class of potential latent variables incorporating the population principal components of the observable covariates. The main reasons are as follows. Theoretically, when latent factors are uncorrelated with each other, factor loadings can generally be obtained through the singular value decomposition (SVD) in factor analysis. Then population principal components would give the unobservable factors. Even if the observed covariates are subject to measurement errors, principal components yield the maximum likelihood estimates of unobservable factors when the measurement errors follow Gaussian distribution with the same variance \citep{Mardia1979}. Practically, principal components evaluate orthogonal directions that reflect maximal variations in the data. In genome-wide association studies, the first few principal components are typically extracted to adjust for population substructures. Similarly, \cite{Leek2007} employed principal components as surrogate variables to estimate the unobservable factors in genome-wide expression studies. Practical examples of such latent variables include the underlying cell types of patients, which can be reflected by the expressions of genes in pathways \citep{Bair2006}.

The major contributions of this paper are threefold. First, we propose nonsparse learning with latent variables based on the aforementioned conditional sparsity structure to simultaneously recover the significant observable predictors and latent factors as well as their effects. By exploring population principal components as common latent variables, it will be helpful in attenuating collinearity and facilitating dimension reduction. Second, to estimate population principal components, we use the sample counterparts and provide the convergence rates of both sample principal components and their score vectors that hold for a wide class of distributions. The convergence property of sample score vectors is critical to the estimation accuracy of latent variables. This is, however, much less studied in the literature compared with the principal components and our work is among the first attempts in the high-dimensional case. Third, we characterize the model identifiability condition and show that the proposed methodology is applicable to general families with properly estimated latent variables. In particular, under some regularity conditions, NSL via the thresholded regression is proved to enjoy model selection consistency and oracle inequalities under various prediction and estimation losses.


The rest of this paper is organized as follows. Section \ref{Sec2} presents the new methodology of nonsparse learning with latent variables. We establish asymptotic properties of sample principal components and their score vectors in high dimensions, as well as theoretical properties of the proposed methodology via the thresholded regression in Section \ref{Sec4}. Simulated and real data examples are provided in Section \ref{Sec6}. Section \ref{Sec7} discusses extensions and possible future work. All the proofs of the main results and additional technical details are included in the Supplementary Material.


\section{Nonsparse learning with latent variables} \label{Sec2}

\subsection{Model setting}

Denote by $\by = (y_1, \dots, y_n)\t$ the $n$-dimensional response vector, $\bX = (\bx_1, \dots, \bx_p)$ the $n \times p$ random design matrix with $p$ predictors, and $\bF = (\bff_1, \dots, \bff_K)$ the $n \times K$ random matrix consisting of $K$ potential latent variables. The latent variables are unobservable but depend on $q$ observable covariates, given by the $n \times q$ random matrix $\bW = (\bw_1, \dots, \bw_q)$. Assume that the rows of $\bX$ and $\bW$ have mean zero and covariance matrices $\bSig_{X}$ and $\bSig_W$, respectively. We consider the following high-dimensional linear regression model with latent variables,
\begin{equation} \label{002}
\by = \bX \bbeta + \bF \bgamma + \bveps,
\end{equation}
where $\bbeta = (\beta_1, \dots, \beta_p)\t$ and $\bgamma = (\gamma_1, \dots, \gamma_K)\t$ are respectively the regression coefficient vectors of observable predictors and latent variables, and $\bveps \sim N(\bzero, \sigma^2 \bI_n)$ is an $n$-dimensional error vector independent of $\bX$ and $\bF$. The conditional sparsity structure is imposed in model (\ref{002}) such that after taking out the impacts of unobservable latent factors $\bF \bgamma$, the true regression coefficient vector of observable predictors $\bbeta_0 = (\beta_{0,1}, \dots, \beta_{0,p})\t$ is sparse. Both the dimensionality $p$ and number of covariates $q$ are allowed to grow nonpolynomially fast with the sample size $n$.

The main difference between model (\ref{002}) and linear regression model in high dimensions lies in the extra latent part $\bF \bgamma$, which accounts for some nonsparse effects of covariates $\bW$. Moreover, when confounding factors potentially exist beyond the original model $\bX \bbeta$, omitting them will cause inconsistency in both variable selection and parameter estimation. See \cite[Chapter 2]{Anderson1980} for detailed discussions of the impacts of confounding variables. Although analyzing the effects of latent variables is challenging as they cannot be measured directly, we will make use of the observable covariates $\bW$ to adjust for confounding as well as the nonsparse effects. Generally speaking, the predictors $\bX$ in model (\ref{002}) stand for features with individual effects while $\bW$ are covariates reflecting the confounding substructures. In practice, which variables should be chosen as $\bX$ and which should be chosen as $\bW$ depend on research interests and the underlying mechanism. Our analysis also allows for a special case that the features $\bX$ and $\bW$ are identical, meaning that the latent factors are nonsparse combinations of the original predictors. The identifiability of this model will be discussed after Condition \ref{robXZ} in Section \ref{Sec4.1}.



Now we illustrate model (\ref{002}) with a practical application. In the analysis of body mass index (BMI), both nutrient intake and gut microbiome composition are believed to be important and they share strong associations \citep{ChenLi2013}. These two groups of features cannot be packed together as predictors in a single model due to their fairly different structures and strong correlations, but we may use nutrient intake as predictors while adjusting for latent factors involving gut microbiome composition. The results of this real data analysis will be presented in Section \ref{Sec6.2}.

As discussed in Section \ref{Sec1}, we focus on one potential family of latent variables incorporating population principal components of the observable covariates $\bW$. Specifically, denote by $\{\bu_i\}_{i = 1}^K$ the top-$K$ principal components (eigenvectors) of the covariance matrix $\Sig_W$, where $K$ is the number of spiked eigenvalues (to be discussed in Section \ref{Sec3nota}) of $\Sig_W$, and is allowed to diverge with the sample size. Then each potential latent variable will be the population principal component score vector, that is, $\bff_i = \bW \bu_i$ for $1 \leq i \leq K$.


\subsection{Estimation procedure by NSL}

With unobservable latent factors $\bF$, it is challenging to consistently estimate and recover the support of the true regression coefficient vector $\bbeta_0$ for observable predictors and the true effects of confounding variables $\bgamma_0 = (\gamma_{0,1}, \dots, \gamma_{0,K})\t$. We partially overcome this difficulty by assuming that the confounding factors appear in an unknown linear form of the observable covariates. Then $\bF$ can be estimated by the sample principal component scores of covariate matrix $\bW$. As $\bW$ has mean zero, the sample covariance matrix $\bS = n^{-1} \bW\t \bW$ is an unbiased estimate of $\bSig_W$ with top-$K$ principal components $\{\widehat{\bu}_i\}_{i = 1}^K$. So the estimated latent variables are $\widehat{\bF} = (\widehat{\bff}_1, \dots, \widehat{\bff}_K)$ with $\widehat{\bff}_i = \bW \widehat{\bu}_i$ for $1 \leq i \leq K$. To ensure model identifiability, both $\bff_i$ and $\widehat{\bff}_i$ are rescaled to have a common $L_2$-norm $n^{1/2}$, matching that of the constant predictor $\bone$ for the intercept. For future prediction, we can transform the coefficient vector $\bgamma$ back by multiplying the scalars $n^{1/2} \|\bW \widehat{\bu}_i\|_2^{-1}$. The notation $\| \cdot \|_q$ denotes the $L_q$-norm of a given vector for $q \in [0, \infty]$.


To produce a joint estimate for the true coefficient vectors $\bbeta_0$ and $\bgamma_0$, we suggest nonsparse learning with latent variables which minimizes
\begin{equation} \label{e001}
Q\left\{(\bbeta\t, \bgamma\t)\t\right\} = (2n)^{-1} \left\|\by - \bX \bbeta - \widehat{\bF} \bgamma\right\|_2^2 + \left\|p_\lambda\left\{(\bbeta_{\ast}\t, \bgamma\t)\t\right\}\right\|_1,
\end{equation}
the penalized residual sum of squares with penalty function $p_\lambda(\cdot)$. Here $\bbeta_{\ast} = (\beta_{\ast,1}, \dots, \beta_{\ast,p})\t$ is the Hadamard (componentwise) product of two $p$-dimensional vectors $\bbeta$ and $(n^{-1/2}\|\bx_k\|_2)_{1 \leq k \leq p}$. It corresponds to the design matrix with each column rescaled to have a common $L_2$-norm $n^{1/2}$. The penalty function $p_\lambda(t)$ is defined on $t \in [0, \infty)$, indexed by $\lambda \geq 0$, and assumed to be increasing in both $\lambda$ and $t$ with $p_\lambda(0) = 0$. We use a compact notation for
\[p_\lambda \left\{(\bbeta_{\ast}\t, \bgamma\t)\t \right\} = \Big\{p_\lambda(|\beta_{\ast,1}|), \dots, p_\lambda(|\beta_{\ast,p}|), p_\lambda(|\gamma_1|), \dots, p_\lambda(|\gamma_K|) \Big\}\t.\]


The proposed methodology in (\ref{e001}) enables the possibility to simultaneously estimate $\bbeta$ and $\bgamma$, identifying the significant observable predictors and latent factors altogether. However, it is still difficult to obtain accurate estimates since the confounding factors $\bF$ are replaced by the estimate $\widehat{\bF}$, and the correlations between the observable predictors and latent variables can aggravate the difficulty. To prevent the estimation errors being further magnified in prediction, we consider $\bgamma$ in an $L_{\infty}$ ball $\mathbb{B}_T = \{\bgamma \in \mathbb{R}^K: \|\bgamma\|_{\infty} \leq T\}$, where any component of $\bgamma$ is assumed to be no larger than $T$ in magnitude. We allow $T$ to diverge slowly such that it will not deteriorate the overall prediction accuracy.



\subsection{Comparisons with existing methods}

The proposed methodology can be regarded as a realization of the aforementioned low rank plus sparse representation \citep{Fan2013} in the linear regression setting, but there are significant differences lying behind them. First, the latent variables in our setup are not necessarily a part of the original predictors, but can stem from any sources related to the underlying features. Second, unlike the typical assumption in factor analysis that the factors and the remaining part are uncorrelated, we allow latent variables to share correlations with the observable predictors. In the extreme case, the latent variables can be nonsparse linear combinations of the predictors. Third, latent variables are employed to recover the information beyond the sparse effects of predictors, and thus we do not modify or assume simplified correlations between the original predictors even after accounting for the latent substructures.

Another method proposed in \cite{Kneip2011} also incorporated principal components as extra predictors in penalized regression, but differs from ours in the following aspects. First of all, based on the framework of factor analysis, the observed predictors in \cite{Kneip2011} were mixtures of individual features and common factors, both of which were unobservable. In view of this, we aim at different scopes of applications. Moreover, \cite{Kneip2011} suggested sparse regression on the projected model, where individual features were recovered as residuals of projecting the observed predictors on the factors. In contrast, we keep the original predictors such that they will not be contaminated when the estimated latent variables are irrelevant. Last but not least, benefitting from factor analysis, the individual features in \cite{Kneip2011} were uncorrelated with each other and also shared no correlation with the factors. But we do not impose such assumptions as explained before.

The proposed methodology is also closely related to principal component regression (PCR). PCR suggests regressing the response vector on a subset of principal components instead of all explanatory variables, and comprehensive properties have been established in the literature for its importance in reducing collinearity and enabling prediction in high dimensions. For instance, \cite{Cook2007} explored the situations where the response can be regressed on the leading principal components of predictors with little loss of information. Probabilistic explanation was provided in \cite{Li2009} to support the phenomenon that the response is often highly correlated with the leading principal components. Our new methodology takes advantage of the strengths of principal components to extract the most relevant information from additional sources and adjust for confounding and nonsparse effects, while the model interpretability is also retained by exploring the individual effects of observable predictors.


\section{Theoretical properties} \label{Sec4} 

We will first establish the convergence properties of sample principal components and their score vectors for a wide class of distributions under the spiked covariance structure. With the aid of them, properties including model selection consistency and oracle inequalities will be proved for the proposed methodology via the thresholded regression using hard-thresholding.


\subsection{Spiked covariance model} \label{Sec3nota}

High-dimensional principal component analysis (PCA) particularly in the context of spiked covariance model, introduced by \cite{Johnstone2001}, has been studied in \cite{Paul2007, Jung2009, Shen2016, Wang2017}, among many others. This model assumes that the first few eigenvalues of the population covariance matrix deviate from one while the rest are equal to one. Although sample principal components are generally inconsistent without strong conditions when the number of covariates is comparable to or larger than the sample size \citep{Johnstone2009}, with the aid of spiked covariance structure, consistency of sample principal components was established in the literature under different high-dimensional settings. For instance, in the high dimension, low sample size context, \cite{Jung2009} proved the consistency of sample principal components for spiked eigenvalues. When both the dimensionality and sample size are diverging, phase transition of sample principal components was studied in \cite{Paul2007, Shen2016} for multivariate Gaussian observations. The asymptotic distributions of spiked principal components were established in \cite{Wang2017} for sub-Gaussian distributions with a finite number of distinguishable spiked eigenvalues.


In this section, we adopt the generalized version of spiked covariance model studied in \cite{Jung2009} for the covariance structure of covariate matrix $\bW$, where the population covariance matrix $\bSig_W$ is assumed to contain $K$ spiked eigenvalues that can be divided into $m$ groups. The eigenvalues grow at the same rate within each group while the orders of magnitude of the $m$ groups are different from each other. To be specific, there are positive constants $\alpha_1 > \alpha_2 > \cdots > \alpha_m > 1$ such that the eigenvalues in the $l$th group grow at the rate of $q^{\alpha_l}$, $1 \leq l \leq m$, where $q$ is the dimensionality or number of covariates in $\bW$. The constants $\alpha_l$ are larger than $1$ since otherwise the sample eigenvectors can be strongly inconsistent \citep{Jung2009}. Denote the group sizes by positive integers $k_1, \dots, k_m$ satisfying $\sum_{l = 1}^m k_l = K < n$. Set $k_{m + 1} = q - K$, which is the number of non-spiked eigenvalues. Then the set of indices for the $l$th group of eigenvalues is
\begin{equation} \label{index}
J_l = \Big\{1 + \sum_{j = 1}^{l - 1}k_j, \dots, k_l + \sum_{j = 1}^{l - 1}k_j\Big\}, \ \ l = 1, \dots, m + 1.
\end{equation}

Although the above eigen-structure looks almost the same as that in \cite{Jung2009}, the key difference lies in the magnitudes of the sample size $n$ and the number of spiked eigenvalues $K$, both of which are allowed to diverge in our setup instead of being fixed. It makes the original convergence analysis of sample eigenvalues and eigenvectors invalid since the number of entries in the dual matrix $\bS_D = n^{-1} \bW \bW\t$ is no longer finite. We will overcome this difficulty by conducting a delicate analysis on the deviation bound of the entries such that the corresponding matrices converge in Frobenius norm. Our theoretical results are applicable to a wide class of distributions including sub-Gaussian distributions. For multivariate Gaussian or sub-Gaussian observations with a finite number of spiked eigenvalues, the phase transition of PCA consistency was studied in, for instance, \cite{Shen2016, Wang2017}. Nevertheless, the convergence property of sample principal component score vectors was not provided in the aforementioned references and needs further investigation.




Assume that the eigen-decomposition of the population covariance matrix $\bSig_W$ is given by $\bSig_W = \bU \bLambda \bU\t$, where $\bLambda$ is a diagonal matrix of eigenvalues $\lambda_1 \geq \lambda_2 \geq \cdots \geq \lambda_q \geq 0$ and $\bU = (\bu_1, \dots, \bu_q)$ is an orthogonal matrix consisting of the population principal components. Analogously, the eigen-decomposition of $\bS = \widehat{\bU}\widehat{\bLambda}\widehat{\bU}\t$ provides the diagonal matrix $\widehat{\bLambda}$ of sample eigenvalues $\widehat{\lambda}_1 \geq \widehat{\lambda}_2 \geq \cdots \geq \widehat{\lambda}_q \geq 0$ and the orthogonal matrix $\widehat{\bU} = (\widehat{\bu}_1, \dots, \widehat{\bu}_q)$ consisting of sample principal components. We always assume that the sample principal components take the correct directions such that the angles between sample and population principal components are no more than a right angle. Our main focus is the high-dimensional setting where the number of covariates $q$ is no less than the sample size $n$. Denote by $\bZ = \bLambda^{-1/2}\bU\t \bW\t$ the sphered data matrix. It is clear that the columns of $\bZ$ are independent and identically distributed (i.i.d.) with mean zero and covariance matrix $\bI_q$. To build our theory, we will impose a tail probability bound on the entry of $\bZ$ and make use of the $n$-dimensional dual matrix $\bS_D = n^{-1} \bZ\t \bLambda \bZ$, which shares the same nonzero eigenvalues with $\bS$. 

\subsection{Thresholded regression using hard-thresholding} \label{Secthred}

As discussed in Section \ref{Sec1}, there are a large spectrum of regularization methods for sparse learning in high dimensions. It has been demonstrated in \cite{FanLv2013} that the popular $L_1$-regularization of Lasso and concave methods can be asymptotically equivalent in thresholded parameter space for polynomially growing dimensionality, meaning that they share the same convergence rates in the oracle inequalities. For exponentially growing dimensionality, concave methods can also be asymptotically equivalent and have faster convergence rates than the Lasso. Therefore, we will show theoretical properties of the proposed methodology via a specific concave regularization method, the thresholded regression using hard-thresholding \citep{Zheng2014}. It utilizes either the hard-thresholding penalty $p_{H, \lambda}(t) = \frac{1}{2} \left[\lambda^2 - (\lambda - t)_+^2\right]$ or the $L_0$-penalty $p_{H_0, \lambda}(t) = 2^{-1} \lambda^2 1_{\{t \neq 0\}}$ in the penalized least squares (\ref{e001}), both of which enjoy the hard-thresholding property \citep[Lemma 1]{Zheng2014} that facilitates sparse modeling and consistent estimation.

A key concept for characterizing model identifiability in \cite{Zheng2014} is the robust spark $\emph{rspark}_c(\bX)$ of a given $n \times p$ design matrix $\bX$ with bound $c$, defined as the smallest possible number $\tau$ such that there exists a submatrix consisting of $\tau$ columns from $n^{-1/2} \tilde{\bX}$ with a singular value less than the given positive constant $c$, where $\tilde{\bX}$ is obtained by rescaling the columns of $\bX$ to have a common $L_2$-norm $n^{1/2}$. The bound on the magnitude of $\emph{rspark}_c(\bX)$ was established in \cite{FanLv2013} for Gaussian design matrices and further studied by \cite{Lv2013} for more general random design matrices. Under mild conditions, $M = \tilde{c} n/(\log p)$ with some positive constant $\tilde{c}$ will provide a lower bound on $\emph{rspark}_c(\bX, \bF)$ for the augmented design matrix (see Condition \ref{robXZ} in Section \ref{Sec4.1} for details). Following \cite{FanLv2013} and \cite{Zheng2014}, we consider the regularized estimator on the union of coordinate subspaces $\mathbb{S}_{M/2} = \{(\bbeta\t, \bgamma\t)\t \in \mathbb{R}^{p + K}: \|(\bbeta\t, \bgamma\t)\t\|_0 < M/2\}$ to ensure model identifiability and reduce estimation instability. So the joint estimator $(\hbbeta\t, \hbgamma\t)\t$ is defined as the global minimizer of the penalized least squares (\ref{e001}) constrained on space $\mathbb{S}_{M/2}$.



\subsection{Technical conditions} \label{Sec4.1}

Here we list a few technical conditions and discuss their relevance. Denote $\Delta = \min_{1 \leq l \leq m - 1} (\alpha_{l} - \alpha_{l + 1})$. Then $q^{\Delta}$ reflects the minimum gap between the magnitudes of spiked eigenvalues in two successive groups. The first two conditions are imposed for Theorem \ref{Thmpred}, while the rest are needed in Theorem \ref{Thm4} to be presented in Section \ref{Sec4.2}.
\begin{condition} \label{eigenvl}
There exist positive constants $c_i$ and $C$ such that uniformly over $i \in J_l$, $1 \leq l \leq m$,
\[\lambda_i/q^{\alpha_l} = c_i + O(q^{-\Delta}) \ \ with \ \ c_i \leq C,\]
and $\lambda_j \leq C$ for any $j \in J_{m + 1}$.
\end{condition}

\begin{condition} \label{tailpb}
(a) There exists some positive $\alpha < \mbox{min}\{\Delta, \alpha_m - 1\}$ such that uniformly over $1 \leq i \leq n$ and $1 \leq j \leq q$, the $(j,i)$th entry of $\bZ$, denoted by $z_{ji}$, satisfies
\begin{align*}
P(z_{ji}^2 > K^{-1} q^{\alpha}) = o(q^{-1} n^{-1}).
\end{align*}
(b) For any $1 \leq l \leq m$, $\|n^{-1} \bZ_l \bZ_l\t - I_{k_l}\|_{\infty} = o(k_l^{-1})$, where $\bZ_l$ is a submatrix of $\bZ$ consisting of the rows with indices in $J_l$.
\end{condition}

\begin{condition} \label{cond3}
Uniformly over $j$, $1 \leq j \leq K$, the angle $\omega_{jj}$ between the $j$th estimated latent vector $\widehat{\bff}_j$ and its population counterpart $\bff_j$ satisfies $\cos (\omega_{jj}) \geq 1 - \frac{c_2^2 \log n}{8 K^2 T^2 n}$ with probability $1 - \theta_1$ that converges to one as $n \to \infty$.
\end{condition}

\begin{condition} \label{robXZ}
The inequality $\|n^{-1/2}(\bX, \bF) \bdelta\|_2 \geq c \|\bdelta\|_2$ holds for any $\bdelta$ satisfying $\|\bdelta\|_0 < M$ with probability $1 - \theta_2$ approaching one as $n \to \infty$.
\end{condition}

\begin{condition} \label{condx}
There exists some positive constant $L$ such that
\[P\Big(\cap_{j = 1}^p \big\{L^{-1} \leq \frac{\|\bx_j\|_2}{\sqrt{n}} \leq L\big\}\Big) = 1 - \theta_3,\]
where $\theta_3$ converges to zero as $n \rightarrow \infty$.
\end{condition}

\begin{condition} \label{cond2}
Denote by $s = \|\bbeta_0\|_0 + \|\bgamma_0\|_0$ and $b_0 = \min_{j \in \supp(\bbeta_0)}(|\beta_{0,j}|) \wedge \min_{j \in \supp(\bgamma_0)}(|\gamma_{0,j}|)$ the number of overall significant predictors and overall minimum signal strength, respectively. It holds that $s < M/2$ and
\[b_0 > [(\sqrt{2}c_1^{-1}) \vee 1] c_1^{-1}c_2 L \sqrt{(2s + 1) (\log p)/n}\]
for some positive constants $c_1$ defined in Proposition \ref{L4} in Section \ref{Sec4.2} and $c_2 > 2\sqrt{2}\sigma$.
\end{condition}

Condition \ref{eigenvl} requires that the orders of magnitude of spiked eigenvalues in each group be the same while their limits can be different, depending on the constants $c_i$. It is weaker than those usually imposed in the literature such as \cite{Shen2016}, where the spiked eigenvalues in each group share exactly the same limit. Nevertheless, we will prove the consistency of spiked sample eigenvalues under very mild conditions. To distinguish the eigenvalues in different groups, convergence to the corresponding limit is assumed to be at a rate of $O(q^{-\Delta})$. As the number of spiked eigenvalues diverges with $q$, we impose a constant upper bound $C$ on $c_i$ for simplicity, and our technical argument still applies when $C$ diverges slowly with $q$. Without loss of generality, the upper bound $C$ also controls the non-spiked eigenvalues.



As pointed out earlier, the columns of the sphered data matrix $\bZ$ are i.i.d. with mean zero and covariance matrix $\bI_p$. Then part (a) of Condition \ref{tailpb} holds as long as the entries in any column of $\bZ$ satisfy the tail probability bound. Moreover, it is clear that this tail bound decays polynomially, so that it holds for a wide class of distributions including sub-Gaussian distributions. With this tail bound, the larger sample eigenvalues would dominate the sum of all eigenvalues in the smaller groups regardless of the randomness. Furthermore, by definition we know that the columns of $\bZ_l$ are i.i.d. with mean zero and covariance matrix $\bI_{k_l}$ such that $n^{-1} \bZ_l \bZ_l\t \to \bI_{k_l}$ entrywise as $n \to \infty$. Hence, part (b) of Condition \ref{tailpb} is a very mild assumption to deal with the possibly diverging group sizes $k_l$.





Condition \ref{cond3} imposes a convergence rate of $\log n/(K^2 T^2 n)$ for the estimation accuracy of confounding factors, so that the estimation errors in $\widehat{\bF}$ will not deteriorate the overall estimation and prediction powers. This rate is easy to satisfy in view of the results in Theorem \ref{Thmpred} in Section \ref{Sec4.2} since the sample principal component score vectors are shown to converge to the population counterparts in polynomial orders of $q$, which is typically larger than $n$ in high-dimensional settings.

Condition \ref{robXZ} assumes the robust spark of matrix $(\bX, \bF)$ with bound $c$ to be at least $M = \tilde{c} n/(\log p)$ with significant probability. It is the key for characterizing the model identifiability in our conditional sparsity structure and also controls the correlations between the observable predictors $\bX$ and latent factors $\bF$. Consider a special case where $\bF$ consists of nonsparse linear combinations of the original predictors $\bX$. Then model (\ref{002}) cannot be identified if we allow for nonsparse regression coefficients. However, if we constrain the model size by certain sparsity level, such as $\emph{rspark}_c(\bX, \bF)$, the model will become identifiable since $\bF$ cannot be represented by sparse linear combinations of $\bX$. Utilizing the same idea, if we impose conditions such as the minimum eigenvalue for the covariance matrix of any $M_1$ features in $(\bX, \bF)$ being bounded from below, where $M_1 = \tilde{c}_1 n/(\log p)$ with $\tilde{c}_1 > \tilde{c}$ denotes the sparsity level, then \cite[Theorem 2]{Lv2013} ensures that the robust spark of any submatrix consisting of less than $M_1$ columns of $(\bX, \bF)$ will be no less than $M = \tilde{c} n/(\log p)$. It holds for general distributions with tail probability decaying exponentially fast with the sample size $n$, and the constant $\tilde{c}$ depending only on $c$. This justifies the inequality in Condition \ref{robXZ}.



While no distributional assumptions are imposed on the random design matrix $\bX$, Condition \ref{condx} puts a mild constraint that the $L_2$-norm of any column vector of $\bX$ divided by its common scale $n^{1/2}$ is bounded with significant probability. It can be satisfied by many distributions and is needed due to the rescaling of $\bbeta_{\ast}$ in (\ref{e001}). Condition \ref{cond2} is similar to that of \cite{Zheng2014} for deriving the global properties via the thresholded regression. The first part puts a sparsity constraint on the true model size $s$ for model identifiability as discussed after Condition \ref{robXZ}, while the second part gives a lower bound $O\{[s (\log p) / n]^{1/2}\}$ on the minimum signal strength to distinguish the significant predictors from the others.

\subsection{Main results} \label{Sec4.2}

We provide two main theorems in this section. The first one is concerned with the asymptotic properties of sample principal components and their score vectors, which serves as a bridge for establishing the global properties in the second theorem.

A sample principal component is said to be consistent with its population counterpart if the angle between them converges to zero asymptotically. However, when several population eigenvalues belong to the same group, the corresponding principal components may not be distinguishable. In that case, subspace consistency is essential to characterizing the asymptotic properties \citep{Jung2009}. Denote $\theta_{il} = Angle(\widehat{\bu}_i, span\{\bu_j: j \in J_l\})$ for $i \in J_l$, $1 \leq l \leq m$, which is the angle between the $i$th sample principal component and the subspace spanned by population principal components in the corresponding spiked group. The following theorem presents the convergence rates of sample principal components in terms of angles under the aforementioned generalized spiked covariance model. Moreover, for the identifiability of latent factors, we assume each group size to be one for the spiked eigenvalues when studying the principal component score vectors. That is, $k_l = 1$ for $1 \leq l \leq m$, implying $K = m$. 

\begin{theorem}[Convergence rates] \label{Thmpred}
Under Conditions \ref{eigenvl} and \ref{tailpb}, with probability approaching one, the following statements hold.

(a) Uniformly over $i \in J_l$, $1 \leq l \leq m$, $\theta_{il} = Angle(\widehat{\bu}_i, span\{\bu_j: j \in J_l\})$ is no more than
\begin{align}\label{convvec}
\arccos([1 - \sum_{t = 1}^{l - 1} \big[\prod_{i = t + 1}^{l - 1} (1 + k_i)\big] O\{k_t A(t)\} - O\{A(l)\}]^{1/2}),
\end{align}
where $A(t) = \big(\sum_{l = t + 1}^m k_l q^{\alpha_l} + k_{m + 1} \big) K^{-1} q^{\alpha - \alpha_t}$ and we define $\sum_{t = i}^{j} s_t = 0$ and $\prod_{t = i}^{j} s_t = 1$ if $j < i$ for any sequence $\{s_t\}$.

\smallskip

(b) If each group of spiked eigenvalues has size one, then uniformly over $1 \leq i \leq K$, $\omega_{ii} = Angle(\bW \widehat{\bu}_i, \bW \bu_i)$ is no more than
\begin{align*}
\arccos([1 - \sum_{t = 1}^{i - 1} 2^{i - t - 1} O\{A(t)\} - O\{A(i)\}]^{1/2}).
\end{align*}
\end{theorem}

Part (a) of Theorem \ref{Thmpred} provides the uniform convergence rates of sample principal components to the corresponding subspaces for general spiked covariance structure with possibly tiered eigenvalues under mild conditions. Since the convergence rates of $\theta^2_{il}$ to zero and $\cos^2(\theta_{il})$ to one are the same by L'Hospital's rule, both of them are $\sum_{t = 1}^{l - 1} \big[\prod_{i = t + 1}^{l - 1} (1 + k_i)\big] O\{k_t A(t)\} + O\{A(l)\}$ in view of (\ref{convvec}). Thus, when the group sizes $k_l$ are relatively small, the convergence rates are determined by $A(t)$, which decays polynomially with $q$ and converges to zero fairly fast. It shows the ``blessing of dimensionality" under the spiked covariance structure since the larger $q$ gives faster convergence rates. Furthermore, it is clear that when the gaps between the magnitudes of different spiked groups are large, $A(t)$ decays quickly with $q$ to accelerate the convergence of sample principal components.

The uniform convergence rates of sample principal component score vectors are given in part (b) of Theorem \ref{Thmpred} when each group contains only one spiked eigenvalue such that the latent factors are separable. In fact, the proof of Theorem \ref{Thmpred} shows that the sample score vectors converge at least as fast as the sample principal components. Then the results in part (b) are essentially the convergence rates in part (a) with $k_l = 1$. Since the number of spiked eigenvalues $K$ is much smaller than $q$, the sample principal component score vectors will converge to the population counterparts polynomially with $q$. The convergence property of sample score vectors is critical to our purpose of nonsparse learning since it offers the estimation accuracy of latent variables, which is much less well studied in the literature. To the best of our knowledge, our work is a first attempt in high dimensions.



The established asymptotic property of sample principal component score vectors justifies the estimation accuracy assumption in Condition \ref{cond3}. Together with Condition \ref{robXZ}, it leads to the following proposition.
\begin{proposition} \label{L4}
Under Conditions \ref{cond3} and \ref{robXZ}, the inequality
\[\|n^{-1/2}(\bX, \widehat{\bF}) \bdelta\|_2 \geq c_1 \|\bdelta\|_2\]
holds for some positive constant $c_1$ and any $\bdelta$ satisfying $\|\bdelta\|_0 < M$ with probability at least $1 - \theta_1 - \theta_2$.
\end{proposition}

From the proof of Proposition \ref{L4}, we see that the constant $c_1$ is smaller than but can be very close to $c$ when $n$ is relatively large. Therefore, Proposition \ref{L4} shows that the robust spark of the augmented design matrix $(\bX, \widehat{\bF})$ will be close to that of $(\bX, \bF)$ when $\bF$ is accurately estimated by $\widehat{\bF}$. We are now ready to present theoretical properties for the proposed methodology.



\begin{theorem}[Global properties] \label{Thm4}
Assume that Conditions \ref{cond3}--\ref{cond2} hold and
\[
c_1^{-1} c_2 \sqrt{(2s + 1) (\log p)/n} < \lambda < L^{-1}b_0 [1 \wedge (c_1 / \sqrt{2})].
\]
Then for both the hard-thresholding penalty $p_{H, \lambda}(t)$ and $L_0$-penalty $p_{H_0, \lambda}(t)$, with probability at least $1 - 4\sigma(2/\pi)^{1/2}c_2^{-1}(\log p)^{-1/2}p^{1 - \frac{c_2^2}{8 \sigma^2}} - 2\sigma(2/\pi)^{1/2} c_2^{-1} s (\log n)^{-1/2} \\ \cdot n^{-\frac{c_2^2}{8 \sigma^2}} - \theta_1 - \theta_2 - \theta_3$, the regularized estimator $(\hbbeta\t, \hbgamma\t)\t$ satisfies that:
\begin{itemize}
\item[\emph{(a)}] \emph{(Model selection consistency)} $\supp \big\{(\hbbeta\t, \hbgamma\t)\t \big\} = \supp \big\{(\bbeta_0\t, \bgamma_0\t)\t \big\}$;

\item[\emph{(b)}] \emph{(Prediction loss)} $n^{-1/2}\|(\bX, \widehat{\bF}) (\hbbeta\t, \hbgamma\t)\t - (\bX, \bF) (\bbeta_0\t, \bgamma_0\t)\t\|_2 \leq (c_2/2 + 2c_2c_1^{-1}\sqrt{s}) \sqrt{(\log n)/n}$;

\item[\emph{(c)}] \emph{(Estimation losses)} $\|\hbbeta - \bbeta_0\|_q \leq 2c_1^{-2}c_2L s^{1/q} \sqrt{(\log n)/n}$, $\|\hbgamma - \bgamma_0\|_q \leq 2c_1^{-2}c_2 s^{1/q} \sqrt{(\log n)/n}$ for $q \in [1, 2]$. The upper bounds with $q = 2$ also hold for $\|\hbbeta - \bbeta_0\|_{\infty}$ and $\|\hbgamma - \bgamma_0\|_{\infty}$.
\end{itemize}
\end{theorem}

The model selection consistency in Theorem \ref{Thm4} shows that we can recover both the significant observable predictors and the latent variables, so that the whole model would be identified by combining these two parts even if it contains nonsparse coefficients. The prediction loss of the joint estimator is shown to be within a logarithmic factor $(\log n)^{1/2}$ of that of the oracle estimator when the regularization parameter $\lambda$ is properly chosen, which is similar to the result in \cite{Zheng2014}. It means that the prediction accuracy is maintained regardless of the hidden effects as long as the latent factors are properly estimated. The extra term $(c_2/2) \sqrt{(\log n)/n}$ in the prediction bound reflects the price we pay in estimating the confounding factors. Furthermore, the oracle inequalities for both $\hbbeta$ and $\hbgamma$ under $L_q$-estimation losses with $q \in [1, 2] \cup \{\infty\}$ are also established in Theorem \ref{Thm4}. Although the estimation accuracy for the nonsparse coefficients $\bU \bgamma$ of $\bW$ are obtainable, we omit the results here since their roles in inferring the individual effects and prediction are equivalent to those of the latent variables. 


The proposed methodology of nonsparse learning with latent variables under the conditional sparsity structure is not restrictive to the potential family of population principal components. It is more broadly applicable to any latent family provided that the estimation accuracy of latent factors in Condition \ref{cond3} and the correlations between the observable predictors and latent factors characterized by the robust spark in Condition \ref{robXZ} hold similarly. The population principal component provides a common and concrete example to extract the latent variables from additional covariates. A significant advantage of this methodology is that even if the estimated latent factors are irrelevant, they rarely affect the variable selection and effect estimation of the original predictors since the number of potential latent variables is generally a small proportion of that of the predictors. This is a key difference between our methodology and those based on factor analysis, which renders it useful for combining additional sources.



\section{Numerical studies} \label{Sec6}
In this section, we discuss the implementation and investigate the finite sample performance of NSL via three regularization methods of the Lasso \citep{Tibshirani1996}, SCAD \citep{Fan2001}, and the thresholded regression using hard-thresholding (Hard) \citep{Zheng2014}. The oracle procedure (Oracle) which knew the true model in advance is also conducted as a benchmark. We will explore two different models, where model $M_1$ involves only observable predictors and model $M_2$ incorporates estimated latent variables as extra predictors. The case of linear regression model (\ref{002}) with the confounding factor as nonsparse combination of the existing predictors is considered in the first example, while in the second example multiple latent factors stem from additional observable covariates and the error vector is relatively heavy-tailed with $t$-distribution.

%

\subsection{Simulation examples} \label{Sec6.1}

\subsubsection{Simulation example 1} \label{Sec6.1.1}

In the first simulation example, we consider a special case of linear regression model (\ref{002}) with potential latent factors $\bF$ coming from the existing observable predictors, that is, $\bW = \bX$. Then $\bF \bgamma$ represents the nonsparse effects of the predictors $\bX$, and it will be interesting to check the impacts of latent variables when they are dense linear combinations of the existing predictors. The sample size $n$ was chosen to be $100$ with true regression coefficient vectors $\bbeta_0 = (\bv\t,\dots,\bv\t,\bzero)\t$, $\bgamma_0 = (0.5,\bzero)\t$, and Gaussian error vector $\bveps \sim N(\textbf{0},\sigma^2 \bI_n)$, where $\bv = (0.6, 0, 0, -0.6, 0, 0)\t$ is repeated $k$ times and $\bgamma_0$ is a $K$-dimensional vector with one nonzero component $0.5$, denoting the effect of the significant confounding factor. We generated $200$ data sets and adopted the setting of $(p, k, K, \sigma) = (1000, 3, 10, 0.4)$ such that there are six nonzero components with magnitude $0.6$ in the true coefficient vector $\bbeta_0$ and ten potential latent variables.

The key point in the design of this simulation study is to construct a population covariance matrix $\bSig$ with spiked structure. Therefore, for each data set, the rows of the $n \times p$ design matrix $\bX$ were sampled as i.i.d. copies from a multivariate normal distribution $N(\bzero, \Sig)$ with $\Sig = \frac{1}{2} (\Sig_1 + \Sig_2)$, where $\Sig_1 = (0.5^{|i-j|})_{1 \leq i, j \leq p}$ and $\Sig_2 = 0.5 \bI_p + 0.5 \bone \bone\t$. The choice of $\Sig_1$ allows for correlation between the predictors at the population level and $\Sig_2$ has an eigen-structure such that the spiked eigenvalue is comparable with $p$. Based on the construction of $\Sig_1$ and $\Sig_2$, it is easy to check that $\Sig$ has the largest eigenvalue $251.75$ and the others are all below $1.75$. For regularization methods, model $M_2$ involved the top-$K$ sample principal components as estimated latent variables while the oracle procedure used the true confounding factor instead of the estimated one. We applied the Lasso, SCAD, and Hard for both $M_1$ and $M_2$ to produce a sequence of sparse models and selected the regularization parameter $\lambda$ by minimizing the prediction error calculated based on an independent validation set for fair comparison of all methods.


To compare the performance of the aforementioned methods under two different models, we consider several performance measures. The first measure is the prediction error (PE) defined as $E (Y - \bx\t \hbbeta)^2$ in model $M_1$ and as $E (Y - \bx\t \hbbeta - \widehat{\bff}\t \hbgamma)^2$ in model $M_2$, where $\hbbeta$ or $(\hbbeta\t, \hbgamma\t)\t$ are the estimated coefficients in the corresponding models, $(\bx\t, Y)$ is an independent test sample of size $10,000$, and $\widehat{\bff}$ is the sample principal component score vector. For the oracle procedure, $\widehat{\bff}$ is replaced by the true confounding factor $\bff$. The second to fourth measures are the $L_q$-estimation losses of $\bbeta_0$, that is, $\|\hbbeta - \bbeta_0\|_q$ with $q = 2, 1$, and $\infty$, respectively. The fifth and sixth measures are the false positives (FP), falsely selected noise predictors, and false negatives (FN), missed true predictors with respect to $\bbeta_0$. We also calculated the estimated error standard deviation $\widehat{\sigma}$ by all methods in both models. The results are summarized in Table \ref{Tab1}. For the selection and effect estimation of latent variables in model $M_2$, we display in Table \ref{Tab2} the measures similar to those defined in Table \ref{Tab1} but with respect to $\bgamma_0$. They are $L_q$-estimation losses $\|\hbgamma - \bgamma_0\|_q$ with $q = 2, 1$, and $\infty$, FP$_{\gamma}$, and FN$_{\gamma}$.

\begin{table}
\caption{\label{Tab1} Means and standard errors (in parentheses) of different performance measures by all methods over 200 simulations in Section \ref{Sec6.1.1}; $M_1$: model with only observable predictors, $M_2$: model includes estimated latent variables} 
\centering
\fbox{
\begin{tabular}{clcccc}
Model         &   Measure               &    Lasso          &     SCAD         &     Hard         &    Oracle \\ 
\hline
$M_1$    &PE                            &    65.27 (1.35)   &     65.29 (1.40) &     68.80 (6.45) &    --- \\    
      &$L_2$-loss                       &    1.61  (0.24)   &     1.61 (0.25)  &     2.25 (1.07)  &    --- \\    
      &$L_1$-loss                       &    4.69  (1.84)   &     4.70 (1.88)  &     5.03 (2.31)  &    --- \\    
      &$L_{\infty}$-loss                &    0.65  (0.13)   &     0.65 (0.15)  &     1.48 (1.10)  &    --- \\    
      &FP                               &    4.45  (7.15)   &     4.45 (7.16)  &     0.51 (0.90)  &    --- \\    
      &FN                               &    5.93  (0.26)   &     5.93 (0.26)  &     5.98 (0.16)  &    --- \\    
      &$\widehat{\sigma}$               &    7.88  (0.57)   &     7.88 (0.57)  &     7.78 (0.60)  &    --- \\    
\hline
$M_2$    &PE                            &    0.39 (0.16)    &     0.19 (0.01)  &    0.19 (0.01)   &    0.17 (0.01) \\ 
      &$L_2$-loss                       &    0.43 (0.13)    &     0.13 (0.03)  &    0.10 (0.03)   &    0.10 (0.03) \\ 
      &$L_1$-loss                       &    1.52 (0.40)    &     0.44 (0.07)  &    0.21 (0.13)   &    0.21 (0.06) \\ 
      &$L_{\infty}$-loss                &    0.23 (0.07)    &     0.07 (0.02)  &    0.07 (0.02)   &    0.07 (0.02) \\ 
      &FP                               &    28.79(6.52)    &     15.99 (5.63) &    0.02 (0.28)   &    0 (0)       \\ 
      &FN                               &    0.02 (0.16)    &     0 (0)        &    0 (0)         &    0 (0)       \\ 
      &$\widehat{\sigma}$               &    0.47 (0.06)    &     0.38 (0.03)  &    0.41 (0.03)   &    0.40 (0.03) \\ 
\end{tabular}}
\end{table}

In view of Table \ref{Tab1}, it is clear that compared with model $M_2$, the performance measures in variable selection, estimation, and prediction all deteriorated seriously in model $M_1$, where most of important predictors were missed and both the estimation and prediction errors were quite large. We want to emphasize that in this first example, the latent variables are linear combinations of the observable predictors initially included in the model, which means that the nonsparse effects would not be captured without the help of estimated confounding factors. On the other hand, the prediction and estimation errors of all regularization methods were reasonably small in the latent variable augmented model $M_2$. And the performance of Hard was comparable to that of the oracle procedure regardless of the estimation errors of latent features, which is in line with the theoretical results in Theorem \ref{Thm4}. Furthermore, we can see from Table \ref{Tab2} that all methods with the estimated latent variables correctly identified the true confounding factor and accurately recovered its effect.

\begin{table}
\caption{\label{Tab2} Means and standard errors (in parentheses) of different performance measures for regression coefficients of confounding factors by all methods over 200 simulations in Section \ref{Sec6.1.1} (The notation $0.00$ denotes a number less than $0.005$.)}
\centering
\fbox{
\begin{tabular}{ccccc}
Measure              &    Lasso         &   SCAD            &     Hard          &    Oracle \\ 
\hline
$L_2$-loss           &    0.02 (0.00)   &    0.01 (0.00)    &     0.01 (0.00)   &    0.00 (0.00) \\ 
$L_1$-loss           &    0.02 (0.01)   &    0.01 (0.00)    &     0.01 (0.00)   &    0.00 (0.00) \\ 
$L_{\infty}$-loss    &    0.02 (0.00)   &    0.01 (0.00)    &     0.01 (0.00)   &    0.00 (0.00) \\ 
FP$_{\gamma}$        &    0.29 (0.55)   &    0.21 (0.43)    &     0 (0)         &    0 (0)       \\ 
FN$_{\gamma}$        &    0 (0)         &    0 (0)          &     0 (0)         &    0 (0)       \\ 
\end{tabular}}
\end{table}

\subsubsection{Simulation example 2} \label{Sec6.1.2}
Now we consider a more general case where the latent variables stem from a group of observable covariates instead of the original predictors. Moreover, we also want to see whether similar results hold when more significant confounding factors are involved and the errors become relatively heavy-tailed. Thus, there are three main changes in the setting of this second example. First, the predictors $\bX$ and observable covariates $\bW$ are different, as well as their covariance structures which will be specified later. Second, there are two significant latent variables and the $K$-dimensional true coefficient vector $\bgamma_0 = (0.5, -0.5, \bzero)\t$. Third, the error vector $\bveps = \sigma \bleta$, where the components of the $n$-dimensional random vector $\bleta$ are independent and follow the $t$-distribution with $df = 10$ degrees of freedom. The settings of $\bbeta_0$ and $(n, p, K, \sigma)$ are the same as in the first simulation example in Section \ref{Sec6.1.1}, while the dimensionality $q$ of covariates $\bW$ equals $1000$, which is also large.

\begin{table}
\caption{\label{Tab3} Means and standard errors (in parentheses) of different performance measures by all methods over 200 simulations in Section \ref{Sec6.1.2}; $M_1$: model with only observable predictors, $M_2$: model includes estimated latent variables, population error standard deviation $\sigma \sqrt{df/(df-2)}$ equals to $0.45$}
\centering
\fbox{
\begin{tabular}{clcccc}
Model    &   Measure          &    Lasso          &     SCAD          &     Hard            &    Oracle \\  
\hline
$M_1$        &PE              &    72.33 (1.53)   &     72.33 (1.53)  &     76.04 (6.62) &    --- \\  
          &$L_2$-loss         &    1.58 (0.24)    &     1.58 (0.24)   &     2.25 (1.09)  &    --- \\  
          &$L_1$-loss         &    4.49 (1.86)    &     4.49 (1.86)   &     5.00 (2.10)  &    --- \\  
          &$L_{\infty}$-loss  &    0.64 (0.13)    &     0.64 (0.13)   &     1.50 (1.15)  &    --- \\  
&FP                           &    3.59 (6.52)    &     3.59 (6.52)   &     0.49 (0.72)  &    --- \\  
&FN                           &    5.95 (0.23)    &     5.95 (0.23)   &     6.00 (0.07)  &    --- \\  
&Error SD                     &    8.33 (0.59)    &     8.33 (0.59)   &     8.20 (0.63)  &    --- \\  
\hline
$M_2$       &PE              &    1.74 (1.08)     &     1.10 (1.05)   &     1.04 (0.99)  &   0.22 (0.01) \\ 
         &$L_2$-loss         &    0.70 (0.22)     &     0.25 (0.22)   &     0.16 (0.18)  &   0.11 (0.03) \\ 
         &$L_1$-loss         &    2.18 (0.50)     &     0.87 (0.50)   &     0.39 (0.53)  &   0.23 (0.07) \\ 
         &$L_{\infty}$-loss  &    0.37 (0.12)     &     0.13 (0.10)   &     0.10 (0.09)  &   0.08 (0.03) \\ 
&FP                          &    20.63 (13.60)   &     23.29 (12.52) &     0.70 (3.34)  &   0 (0) \\       
&FN                          &    0.09 (0.38)     &     0.15 (0.94)   &     0.09 (0.63)  &   0 (0) \\       
&Error SD                    &    0.74 (0.20)     &     0.48 (0.21)   &     0.50 (0.12)  &   0.45 (0.04) \\ 
\end{tabular}}
\end{table}

For the covariance structure of $\bX$, we set $\Sig_X = (0.5^{|i-j|})_{1 \leq i, j \leq p}$ to allow for correlation at the population level. On the other hand, in order to estimate the principal components in high dimensions, the population covariance matrix of $\bW$ should have multiple spiked eigenvalues. Thus, we constructed it using the block diagonal structure such that
\[\Sig_W = \left(\begin{array}{cc} \Sig_{11} & \bzero \\ \bzero & \Sig_{22} \end{array} \right),\]
where $\Sig_{11} = \frac{3}{4} (\Sig_1 + \Sig_2)_{1 \leq i,j \leq 200}$ and $\Sig_{22} = \frac{1}{2} (\Sig_1 + \Sig_2)_{1 \leq i,j \leq 800}$ with the defintions of $\Sig_1$ and $\Sig_2$ similar to those in Section \ref{Sec6.1.1} except for different dimensions. Under such construction, the two largest eigenvalues of $\Sig_W$ are $201.75$ and $77.61$, respectively, while the others are less than $2.63$. Based on the aforementioned covariance structures, for each data set, the rows of $\bX$ and $\bW$ were sampled as i.i.d. copies from the corresponding multivariate normal distribution.

We included the top-$K$ sample principal components in model $M_2$ as potential latent factors and compared the performance of the Lasso, SCAD, Hard, and Oracle by the same performance measures as defined in Section \ref{Sec6.1.1}. The results are summarized in Tables \ref{Tab3} and \ref{Tab4}. From Table \ref{Tab3}, it is clear that the methods which replied only on the observable predictors still suffered a lot under this more difficult setting, where all true predictors were missed, prediction errors were large, and the error standard deviation (SD) was poorly estimated. In contrast, the new NSL methodology via the Lasso, SCAD, and Hard was able to tackle the issues associated with variable selection, coefficient estimation, prediction, and error SD estimation. With the latent variable augmented model $M_2$, Hard almost recovered the exact underlying model. Similar to the first example, in view of Table \ref{Tab4}, all methods correctly identified the significant confounding factors and estimated their effects accurately. However, compared with Tables  \ref{Tab1} and  \ref{Tab2}, most of the performance measures deteriorated in this second example. This is mainly due to the relatively heavy-tailed random errors, as well as the difficulty in estimating multiple high-dimensional principal components.

\begin{table}
\caption{\label{Tab4} Means and standard errors (in parentheses) of different performance measures for regression coefficients of confounding factors by all methods over 200 simulations in Section \ref{Sec6.1.2} (The notation $0.00$ denotes a number less than $0.005$.)}
\centering
\fbox{
\begin{tabular}{ccccc}
Measure                 &    Lasso         &     SCAD         &    Hard            &    Oracle      \\ 
\hline
$L_2$-loss              &    0.08 (0.03)   &     0.07 (0.04)  &    0.07 (0.04)     &    0.01 (0.00) \\ 
$L_1$-loss              &    0.10 (0.04)   &     0.09 (0.05)  &    0.08 (0.05)     &    0.01 (0.00) \\ 
$L_{\infty}$-loss       &    0.08 (0.03)   &     0.06 (0.03)  &    0.06 (0.03)     &    0.01 (0.00) \\ 
FP$_{\gamma}$           &    0.21 (0.45)   &     0.34 (0.60)  &    0.01 (0.10)     &    0 (0) \\       
FN$_{\gamma}$           &    0 (0)         &     0 (0)        &    0 (0)           &    0 (0) \\       
\end{tabular}}
\end{table}

\subsection{Application to nutrient intake with gut microbiome data} \label{Sec6.2}

Nutrient intake strongly affects human health or diseases such as obesity, while gut microbiome composition is an important factor in energy extraction from the diet. We illustrate the usefulness
of our proposed methodology by applying it to the data set reported in \cite{Wu2011} and previously studied by \cite{ChenLi2013} and \cite{Lin2014}, where a cross-sectional study of $98$ healthy volunteers was carried out to investigate the habitual diet effect on the human gut microbiome. The nutrient intake consisted of $214$ micronutrients collected from the volunteers by a food frequency questionnaire. The values were normalized by the residual method to adjust for caloric intake and then standardized to have mean zero and standard deviation one. Similar to \cite{ChenLi2013}, we used one representative for a set of highly correlated micronutrients whose correlation coefficients are larger than $0.9$, resulting in $119$ representative micronutrients in total. Furthermore, stool samples were collected and DNA samples were analyzed by 454/Roche pyrosequencing of 16S rDNA gene segments from the V1--V2 region. After taxonomic assignment of the denoised pyrosequences, the operational taxonomic units were combined into $87$ genera which appeared in at least one sample. We are interested in identifying the important micronutrients and potential latent factors from the gut microbiome genera that are associated with the body mass index (BMI).

Due to the high correlations between the micronutrients, we applied NSL via the elastic net \citep{Zou2005} to this data set by treating BMI, nutrient intake, and gut microbiome composition (after the centered log-ratio transformation \citep{Aitchison83}) as the response, predictors, and covariates of confounding factors, respectively. The data set was split $100$ times into a training set of $60$ samples and a validation set of the remaining samples. For each splitting of the data set, we explored two different models $M_1$ and $M_2$ as defined in Section \ref{Sec6.1} with the top-$20$ sample principal components (PCs) of gut microbiome composition included in model $M_2$ to estimate the potential latent factors. All predictors were rescaled to have a common $L_2$-norm of $n^{1/2}$ and the tuning parameter was chosen by minimizing the prediction error calculated on the validation set. We summarize in Table \ref{Tab5} the selection probabilities and coefficients of the significant micronutrients and latent variables whose selection probabilities were above $0.9$ in $M_1$ or above $0.85$ in $M_2$. The means (with standard errors in parentheses) of the prediction errors averaged over $100$ random splittings were $167.9 \ (7.2)$ in model $M_1$ and $110.3 \ (4.0)$ in model $M_2$, while the median model size also reduced from $93$ to $69$ after applying the NSL methodology. It shows that the prediction performance was improved after utilizing the information of gut microbiome genera.

In view of the model selection results in Table \ref{Tab5}, many significant micronutrients in model $M_1$ became insignificant after adjusting for the latent substructures, which implies that either they affect BMI through the gut microbiome genera or their combinative effects are captured by the latent variables. This was also evidenced by the reduction in the model size mentioned before. Moreover, the effects of some micronutrients changed signs in model $M_2$ and the subsequent associations with BMI are consistent with scientific discoveries \citep{Gul17}. For instance, aspartame is a sugar substitute widely used in beverages such as the diet coke, and it was negatively associated with BMI in model $M_1$ but tended to share a positive association after accounting for the gut microbiome genera. A potential reason is that the people who drink diet coke can have a relatively healthy habitual diet and gut microbiome composition which in turn lower the BMI, but the diet coke itself does not reduce fats. Similar phenomena happened to acrylamide and vitamin $E$ as well.

We also applied the model-free knockoffs \citep{Candes16} with the target FDR level $0.2$ on model $M_2$, and the most significant predictors identified were the latent variables of 7th and 9th PCs. The major gut microbiome genera in the compositions of these two latent variables are displayed in Table \ref{Tab6}. At the phylum level, the latent factors mainly consist of bacteroidetes and firmicutes, whose relative proportion has been shown to affect human obesity \citep{Ley2006}. In view of the associations with BMI, both the 7th and 9th PCs confirm the claim that firmicutes-enriched microbiome holds a greater metabolic potential for energy gain from the diet which results in the gain of weight \citep{Turnbaugh2006}. Furthermore, one of the major microbiome genera in the latent factor of 9th PC, Acidaminococcus, was also found to be positively associated with the BMI in \cite{Lin2014}, which shows that human obesity can be affected at the genus level.


\begin{table}
\caption{\label{Tab5} Selection probabilities and rescaled coefficients (in parentheses) of the most frequently selected predictors by each model across 100 random splittings in Section \ref{Sec6.2}; $M_1$: model with only micronutrients as predictors, $M_2$: model includes latent variables from gut microbiome composition}
\centering
\fbox{%
\scalebox{0.93}{
\begin{tabular}{llllll}
Predictor              & Model $M_1$           & Model $M_2$           & Predictor           & Model $M_1$          & Model $M_2$   \\
\hline
Sodium                 & 0.98 (1.35)           & 0.67 (0.55)           &  PC($7$th)          & ---------            & 0.99 (1.76)   \\
Eicosenoic acid        & 0.98 (-2.47)          & 0.80 (-1.24)          &  PC($6$th)          & ---------            & 0.96 (-1.21)  \\
Vitamin $B_{12}$       & 0.96 (0.43)           & 0.62 (0.30)           &  Apigenin           & 0.95 (-1.67)         & 0.93 (-1.88)  \\
Gallocatechin          & 0.96 (-4.81)          & 0.84 (-1.70)          &  PC($9$th)          & ---------            & 0.88 (-0.87)  \\
Riboflavin pills       & 0.94 (1.71)           & 0.55 (0.61)           &  PC($10$th)         & ---------            & 0.86 (0.78)   \\
Acrylamide             & 0.94 (-0.34)          & 0.62 (0.32)           &  Iron               & 0.93 (1.22)          & 0.86 (0.75)   \\
Naringenin             & 0.94 (1.11)           & 0.58 (0.32)           &  Aspartame          & 0.93 (-0.46)         & 0.79 (0.59)   \\
Pelargonidin           & 0.94 (-1.15)          & 0.75 (-1.03)          &  Vitamin $C$        & 0.93 (-0.71)         & 0.76 (-0.39)  \\
Lauric acid            & 0.93 (1.88)           & 0.71 (0.50)           &  Vitamin $E$        & 0.92 (0.45)          & 0.65 (-0.29)  \\
\end{tabular}}}
\end{table}

\begin{table}
\caption{\label{Tab6} Major gut microbiome genera in the compositions of the two significant latent variables identified by the model-free knockoffs in Section \ref{Sec6.2}}
\centering
\fbox{%
\scalebox{1.0}{
\begin{tabular}{clll}
Latent variable     & Phylum            & \quad Genus                     &   Weight       \\
\hline
PC($7$th)           & Firmicutes        & \quad \emph{Dialister}          &   -0.40       \\
                    & Firmicutes        & \quad \emph{Eubacterium}        &  \ 0.39       \\
                    & Bacteroidetes     & \quad \emph{Barnesiella}        &   -0.28       \\
\hline
PC($9$th)           & Firmicutes        & \quad \emph{Acidaminococcus}    &   -0.51       \\
                    & Firmicutes        & \quad \emph{Megasphaera }       &   -0.36       \\
                    & Firmicutes        & \quad \emph{Ruminococcus}       &   -0.30       \\
\end{tabular}}}
\end{table}


\section{Discussions} \label{Sec7}
In this paper, we have introduced a new methodology NSL for prediction and variable selection in the presence of nonsparse coefficient vectors through the conditional sparsity structure, where latent variables are exploited to capture the nonsparse combinations of either the original predictors or additional covariates. The suggested methodology is ideal for the applications including two sets of predictors that cannot be packed directly for analysis, as in our BMI study. Both theoretical guarantees and empirical performance of the potential latent family incorporating population principal components have been demonstrated. And our methodology is also applicable to more general families with properly estimated latent variables and identifiable models.



It would be interesting to further investigate several problems such as hypothesis testing and false discovery rate control in nonsparse learning by the idea of NSL. Based on the established model identifiability condition which characterizes the correlations between observable and latent predictors, hypothesis testing can be proceeded using the de-biasing idea in \cite{Java14, Geer14, Zhang2014}, and false discovery rate could be controlled by applying the knockoffs inference procedures \citep{Barber15, Candes16, Fan17} on the latent variable augmented model. The main difficulty lies in analyzing how the estimation errors of unobservable factors affect the corresponding procedures. Another possible direction is to explore more general ways of modeling the latent variables to deal with the nonsparse coefficient vectors. These problems are beyond the scope of the current paper and will be interesting topics for future research.

\appendix


\bibliographystyle{unsrt}

\newpage

\setcounter{page}{1}
\setcounter{section}{0}
\setcounter{equation}{0}

\renewcommand{\theequation}{A.\arabic{equation}}
\setcounter{equation}{0}

\begin{center}{\bf \large Supplementary Material to ``Nonsparse learning with latent variables''}

\bigskip

Zemin Zheng, Jinchi Lv and Wei Lin
\end{center}

\noindent This Supplementary Material consists of two parts. Section \ref{SecB} lists the key lemmas and presents the proofs for main results. Additional technical proofs for the lemmas are provided in Section \ref{SecL}.

\label{sec:app}

\section{Proofs of main results} \label{SecB}

\subsection{Lemmas} \label{Seclem}

\smallskip

The following lemmas are used in the proofs of main results.
\begin{lemma}[Consistency of spiked sample eigenvalues] \label{Thm1}
Under Conditions \ref{eigenvl} and \ref{tailpb}, with asymptotic probability one, the eigenvalues of the sample covariance matrix $\bS$ satisfy that for any $l$, $1 \leq l \leq m$, uniformly over $i \in J_l$,
\[q^{-\alpha_l}\widehat{\lambda}_i \to c_i \ \text{as} \ q \to \infty.\]
\end{lemma}

\begin{lemma} \label{L3}
Denote by $\bX_0$ and $\widehat{\bF}_0$ the submatrices of $\bX$ and $\widehat{\bF}$ consisting of columns in $\supp(\bbeta_0)$ and $\supp(\bgamma_0)$, respectively, and $\widetilde{\bveps} = (\bF - \widehat{\bF}) \bgamma + \bveps$. For the following two events
\begin{align*}
\widetilde{\mathcal{E}} &= \left\{\|n^{-1} (\bX, \widehat{\bF})\t \widetilde{\bveps}\|_{\infty} \leq c_2 \sqrt{(\log p)/n} \right\} \ \text{ and } \ \\
\widetilde{\mathcal{E}}_0 &= \left\{\|n^{-1} (\bX_0, \widehat{\bF}_0)\t \widetilde{\bveps}\|_{\infty} \leq c_2 \sqrt{(\log n)/n}\right\}
\end{align*}
with constant $c_2 > 2\sqrt{2}\sigma$, when the estimation error bound of $\widehat{\bF}$ in Condition \ref{cond3} holds and the columns of $\bX$ adopt a common scale of $L_2$-norm $n^{1/2}$, we have
\begin{align*}
P(\widetilde{\mathcal{E}} \cap \widetilde{\mathcal{E}}_0) \geq 1 - \frac{4\sqrt{2}\sigma}{c_2 \sqrt{\pi \log p}} p^{1 - \frac{c_2^2}{8 \sigma^2}} - \frac{2\sqrt{2}\sigma s}{c_2 \sqrt{\pi \log n}} n^{-\frac{c_2^2}{8 \sigma^2}},
\end{align*}
which converges to one as $n \to \infty$.
\end{lemma}

\subsection{Proof of Theorem \ref{Thmpred}}

\smallskip

\textbf{Proof of part (a).} In this part, we will focus on the convergence rates of the sample eigenvectors. The key ingredient of this proof is to link the angle between the sample eigenvector and the space spanned by population eigenvectors with the sum of inner products between the sample and population eigenvectors by the $\cos(\cdot)$ function. In this way, it suffices to show that the sum of inner products converges to one for subspace consistency, and at the same time, deriving the convergence rates by induction. To ease readability, we will finish the proof in four steps.

\medskip

\noindent \textbf{Step 1: Analysis of the subspace consistency}. We first show that for any $i \in J_l$, $1 \leq l \leq m$, the subspace consistency of the sample eigenvector $\widehat{\bu}_i$ is equivalent to
\begin{equation} \label{innerpro}
\sum_{j \in J_l} p_{ji}^2 \to 1,
\end{equation}
where $p_{ji} = \bu_j\t\widehat{\bu}_i$ is the inner product between the population eigenvector $\bu_j$ (the $j$th column of $\bU$) and $\widehat{\bu}_i$ (the $i$th column of $\widehat{\bU}$).

Since $\bU$ and $\widehat{\bU}$ are obtained through eigen-decomposition, we know that $\|\bu_j\|_2 = 1$ and $\|\widehat{\bu}_i\|_2 = 1$ for any $i$ and $j$, $1 \leq i,j \leq q$. Note that $\sum_{j \in J_l} (\bu_j\t\widehat{\bu}_i)\bu_j$ is the projection of $\widehat{\bu}_i$ on the space $span\{\bu_j: j \in J_l\}$. It gives
\begin{align*}
& Angle(\widehat{\bu}_i, span\{\bu_j: j \in J_l\}) = \arccos \Big\{\frac{\widehat{\bu}_i\t[\sum_{j \in J_l} (\bu_j\t\widehat{\bu}_i)\bu_j]}{\|\widehat{\bu}_i\|_2 \cdot \|\sum_{j \in J_l} (\bu_j\t\widehat{\bu}_i)\bu_j\|_2}\Big\} = \\
&\arccos\Big\{\frac{\sum_{j \in J_l} (\bu_j\t\widehat{\bu}_i)^2}{[\sum_{j \in J_l} (\bu_j\t\widehat{\bu}_i)^2]^{1/2}}\Big\} = \arccos \big\{\sqrt{\sum_{j \in J_l} (\bu_j\t\widehat{\bu}_i)^2}\big\} = \arccos\big\{\big(\sum_{j \in J_l} p_{ji}^2\big)^{1/2}\big\}.
\end{align*}
Thus, $Angle(\widehat{\bu}_i, span\{\bu_j: j \in J_l\}) \to 0$ is equivalent to $\sum_{j \in J_l} p_{ji}^2 \to 1$ as $q \to \infty$ for any $i \in J_l$, $1 \leq l \leq m$. Moreover, the convergence rate of $\sum_{j \in J_l} p_{ji}^2$ indeed provides the convergence rate of the sample eigenvector $\widehat{\bu}_i$ to the corresponding space of population eigenvectors.

We will then prove the convergence rates by induction. Hereafter our analysis will be conditional on the event $\mathcal{E}$, which is defined in the proof of Lemma \ref{Thm1} for the consistency of the spiked sample eigenvalues and enjoys asymptotic probability one.

\medskip

\noindent \textbf{Step 2: Convergence rates of sample eigenvectors with indices in $J_1$}. This step aims at proving that uniformly over $i \in J_1$, the convergence rate of $\sum_{j \in J_1} p_{ji}^2$ is given by
\begin{equation} \label{At}
\sum_{j \in J_1} p_{ji}^2 \geq 1 - O\{\big(\sum_{l = 2}^m k_l q^{\alpha_l} + k_{m + 1} \big) K^{-1} q^{\alpha - \alpha_1}\} = 1 - O\{A(1)\},
\end{equation}
where $A(t) = \big(\sum_{l = t + 1}^m k_l q^{\alpha_l} + k_{m + 1} \big) K^{-1} q^{\alpha - \alpha_t}$ is defined in Theorem \ref{Thmpred}. It is also the first part of induction. Let $\bP = \bU\t \widehat{\bU} = \{p_{ij}\}_{1 \leq i,j \leq q}$. We have $\sum_{j = 1}^q p_{ji}^2 = 1$ for any $i$ since $\bP$ is a unitary matrix. To prove (\ref{At}), it suffices to show $$\sum_{j \in J_2 \cup \cdots \cup J_{m + 1}} p_{ji}^2 \leq O\{A(1)\}.$$

Recall that $\bZ = \bLambda^{-1/2}\bU\t \bW\t$, $\bS = n^{-1} \bW\t \bW  = \widehat{\bU}\widehat{\bLambda}\widehat{\bU}\t$. Therefore, we get a connection between $\bZ$ and $\bP$ that
\[n^{-1} \bZ \bZ\t = n^{-1} \bLambda^{-1/2} \bU\t \bW\t \bW \bU \bLambda^{-1/2} = \bLambda^{-1/2} \bP \widehat{\bLambda} \bP\t \bLambda^{-1/2}.\]
For any $j$, $1 \leq j \leq q$, in view of the $(j,j)$th entry, the above equality gives
\begin{equation} \label{pij}
\lambda_j^{-1}\sum_{i = 1}^q \widehat{\lambda}_i p_{ji}^2 = n^{-1}\bz_j\t \bz_j,
\end{equation}
where $\bz_j$ is the $j$th column vector of $\bZ\t$. It implies for any $i$, $1 \leq i \leq q$, $\lambda_j^{-1} \widehat{\lambda}_i p_{ji}^2 \leq n^{-1}\bz_j\t \bz_j$. Based on this fact, we have
\begin{equation} \label{equ1}
\sum_{j \in J_2 \cup \cdots \cup J_{m + 1}} p_{ji}^2 \leq \sum_{j \in J_2 \cup \cdots \cup J_{m + 1}} n^{-1}\bz_j\t \bz_j \lambda_j / \widehat{\lambda}_i = \sum_{t = 1}^n \sum_{j \in J_2 \cup \cdots \cup J_{m + 1}} z_{jt}^2 \lambda_j / (n \widehat{\lambda}_i),
\end{equation}
where $z_{jt}$ is the $(j,t)$th entry of $\bZ$. Conditional on the event $\mathcal{E}$, by Lemma \ref{Thm1}, Conditions \ref{eigenvl} and \ref{tailpb}, we have
\begin{align} \label{align1}
&\sum_{t = 1}^n  \sum_{j \in J_2 \cup \cdots \cup J_{m + 1}} z_{jt}^2 \lambda_j / (n \widehat{\lambda}_i) \leq \sum_{j \in J_2 \cup \cdots \cup J_{m + 1}} K^{-1} q^{\alpha} \lambda_j / \widehat{\lambda}_i \nonumber\\
= & \ O\{K^{-1} q^{\alpha} C \big(\sum_{l = 2}^m k_l q^{\alpha_l} + k_{m + 1}\big)/q^{\alpha_1}\} = O\{A(1)\}.
\end{align}

Since the convergences of $\widehat{\lambda}_i$ are uniform over $i \in J_1$ by Lemma \ref{Thm1}, the above inequality holds uniformly over $i \in J_1$. Inequalities (\ref{equ1}) and (\ref{align1}) together entail $\sum_{j \in J_2 \cup \cdots \cup J_{m + 1}} p_{ji}^2 \leq O\{A(1)\}$ uniformly over $i \in J_1$, which implies the convergence rate in (\ref{At}) for the sample eigenvectors with indices in $J_1$. It shows that when $s = 1$, the convergence rate coincides with our claim that uniformly over $i \in J_{l}$, $1 \leq l \leq s$,
\begin{align}\label{claim}
\sum_{j \in J_{l}} p_{ji}^2 \geq 1 - \sum_{t = 1}^{l - 1} \big[\prod_{i = t + 1}^{l - 1} (1 + k_i)\big] O\{k_t A(t)\} - O\{A(l)\}.
\end{align}
Note that we define $\sum_{t = a}^{b} s_t = 0$ and $\prod_{t = a}^{b} s_t = 1$ if $b < a$ for any positive sequence $\{s_t\}$.

\medskip

\noindent \textbf{Step 3: Convergence rates of sample eigenvectors with indices in $J_2$}. Before formally completing the proof by induction, we would like to derive the convergence rates of $\sum_{j \in J_2} p_{ji}^2$ directly for $i \in J_2$ to get the basic idea of induction.

Since we already proved the convergence rate in (\ref{At}) uniformly over $i \in J_1$ in \textbf{Step 2}, summing over $i \in J_1$ gives
\begin{align}\label{key1}
\sum_{i \in J_1} \sum_{j \in J_1} p_{ji}^2 \geq k_1(1 - O\{A(1)\}) = k_1 - O\{k_1 A(1)\}.
\end{align}
Along with the fact that $\sum_{i = 1}^q p_{ji}^2 = 1$, we get
\begin{align}\label{key2}
\sum_{i \in J_2 \cup \cdots \cup J_{m + 1}} & \sum_{j \in J_1} p_{ji}^2 = \sum_{i = 1}^q \sum_{j \in J_1} p_{ji}^2 - \sum_{i \in J_1} \sum_{j \in J_1} p_{ji}^2  = \sum_{j \in J_1} \sum_{i = 1}^q p_{ji}^2 - \sum_{i \in J_1} \sum_{j \in J_1} p_{ji}^2 \nonumber \\
= k_1 - & \ \sum_{i \in J_1} \sum_{j \in J_1} p_{ji}^2 \leq k_1 - (k_1 - O\{k_1 A(1)\}) = O\{k_1 A(1)\}.
\end{align}
The above result is important as it also implies that uniformly over $i \in J_2$,
\begin{align}\label{k1A}
\sum_{j \in J_1} p_{ji}^2 \leq O\{k_1 A(1)\}.
\end{align}

For the sample eigenvector $\widehat{u}_i$ with index $i \in J_2$, in order to find a lower bound for $\sum_{j \in J_2} p_{ji}^2$, we write it as
\begin{equation}\label{thetai2}
\sum_{j \in J_2} p_{ji}^2 = 1 \ - \ \sum_{j \in J_1} p_{ji}^2 \ - \sum_{j \in J_3 \cup \cdots \cup J_{m + 1}} p_{ji}^2.
\end{equation}
The upper bound of $\sum_{j \in J_1} p_{ji}^2$ was provided in (\ref{k1A}). For the second term $\sum_{j \in J_{3} \cup \cdots \cup J_{m + 1}} p_{ji}^2$, similar to (\ref{equ1}) and (\ref{align1}) in \textbf{Step 2}, by Lemma \ref{Thm1}, Conditions \ref{eigenvl} and \ref{tailpb}, we have uniformly over $i \in J_2$,
\begin{align*}
\sum_{j \in J_{3} \cup \cdots \cup J_{m + 1}} p_{ji}^2 \leq O\{K^{-1} q^{\alpha} C \big(\sum_{l = 3}^m k_l q^{\alpha_l} + k_{m + 1} \big)/q^{\alpha_2}\} = O\{A(2)\}.
\end{align*}
Plugging the above two bounds into (\ref{thetai2}) gives
\begin{align*}
\sum_{j \in J_2} p_{ji}^2 \geq 1 - O\{k_1 A(1)\} - O\{A(2)\},
\end{align*}
which shows that the uniform convergence rate of the sample eigenvectors $\widehat{u}_i$ over $i \in J_2$. Together with the uniform convergence rate over $i \in J_1$ established in \textbf{Step 2}, our claim in (\ref{claim}) gives the uniform convergence rates of the sample eigenvectors $\widehat{u}_i$ over $i \in J_1 \cup J_2$.

\medskip

\noindent \textbf{Step 4: Convergence rates of sample eigenvectors with indices in $J_3$ to $J_m$}. In this step, we will complete the proof by induction. Specifically, we show that the claim in (\ref{claim}) holds for any fixed $s$, $3 \leq s \leq m$, based on the induction assumption that the claim holds for $s - 1$.

By the induction assumption, we have uniformly over $i \in J_l$, $1 \leq l \leq s - 1$,
\begin{align*}
\sum_{j \in J_{l}} p_{ji}^2 & \geq 1 - \sum_{t = 1}^{l - 1} \big[\prod_{i = t + 1}^{l - 1} (1 + k_i)\big] O\{k_t A(t)\} - O\{A(l)\}.
\end{align*}
By a similar argument as in (\ref{key1}) and (\ref{key2}), it follows that
\begin{align}\label{Js}
\sum_{i \in J_{l + 1} \cup \cdots \cup J_{m + 1}} \sum_{j \in J_{l}} p_{ji}^2 \leq k_l \big(\sum_{t = 1}^{l - 1} \big[\prod_{i = t + 1}^{l - 1} (1 + k_i)\big] O\{k_t A(t)\} + O\{A(l)\}\big).
\end{align}
Similarly as in \textbf{Step 3}, for any $i \in J_{s}$, to get the convergence rate of $\sum_{j \in J_{s}} p_{ji}^2$, we write it as
\begin{align*}
\sum_{j \in J_{s}} p_{ji}^2 = 1 \ \ - \sum_{j \in J_1 \cup \cdots \cup J_{s - 1}} p_{ji}^2 \ \ - \sum_{j \in J_{s + 1} \cup \cdots \cup J_{m + 1}} p_{ji}^2.
\end{align*}

We will first derive the convergence rate of $\sum_{j \in J_1 \cup \cdots \cup J_{s - 1}} p_{ji}^2$. When $1 \leq l \leq s - 1$, we have $i \in J_{s} \subset J_{l + 1} \cup \cdots \cup J_{m + 1}$. In view of (\ref{Js}), it gives that uniformly over $i \in J_{s}$,
\begin{align*}
\sum_{j \in J_{l}} p_{ji}^2 \leq k_l\big(\sum_{t = 1}^{l - 1} \big[\prod_{i = t + 1}^{l - 1} (1 + k_i)] O\{k_t A(t)\} + O\{A(l)\}\big).
\end{align*}
Summing over $l = 1, \dots, s - 1$, we get
\begin{align*}
\sum_{j \in J_1 \cup \cdots \cup J_{s - 1}} p_{ji}^2 \leq \sum_{l = 1}^{s - 1} k_l \big(\sum_{t = 1}^{l - 1} \big[\prod_{i = t + 1}^{l - 1} (1 + k_i)\big] O\{k_t A(t)\} + O\{A(l)\}\big).
\end{align*}
To simplify the above expression, exchanging the summation order with respect to $l$ and $t$ gives
\begin{align*}
\sum_{j \in J_1 \cup \cdots \cup J_{s - 1}} & p_{ji}^2 \leq \sum_{l = 1}^{s - 1} \sum_{t = 1}^{l - 1} k_l \big[\prod_{i = t + 1}^{l - 1}(1 + k_i)\big] O\{k_t A(t)\} + \sum_{l = 1}^{s - 1} O\{k_l A(l)\}\\
& = \sum_{t = 1}^{s - 2} \sum_{l = t + 1}^{s - 1} k_l \big[\prod_{i = t + 1}^{l - 1}(1 + k_i)\big] O\{k_t A(t)\} + \sum_{t = 1}^{s - 1} O\{k_t A(t)\}.
\end{align*}

Then we combine the coefficients of $k_t A(t)$ to get
\begin{align*}
\sum_{j \in J_1 \cup \cdots \cup J_{s - 1}} p_{ji}^2 \leq \sum_{t = 1}^{s - 2} O\{k_t A(t)\} \big(1 + \sum_{l = t + 1}^{s - 1} k_l \big[\prod_{i = t + 1}^{l - 1}(1 + k_i)\big] \big) + \sum_{t = s - 1} O\{k_t A(t)\}.
\end{align*}
Since it is immediate to conclude by induction that
\begin{align*}
1 + &\sum_{l = t + 1}^{s - 1} k_l \big[\prod_{i = t + 1}^{l - 1}(1 + k_i)\big] = 1 + k_{t + 1} + k_{t + 2}(1 + k_{t + 1}) + \cdots \\ 
&+ k_{s - 1} (1 + k_{s - 2})(1 + k_{s - 3})\cdots(1 + k_{t + 2})(1 + k_{t + 1}) = \prod_{i = t + 1}^{s - 1} (1 + k_i),
\end{align*}
we then have
\begin{align*}
\sum_{j \in J_1 \cup \cdots \cup J_{s - 1}} p_{ji}^2 &\leq \sum_{t = 1}^{s - 2} \big[\prod_{i = t + 1}^{s - 1}(1 + k_i)\big] O\{k_t A(t)\} + \sum_{t = s - 1} O\{k_t A(t)\} \\
&= \sum_{t = 1}^{s - 1} \big[\prod_{i = t + 1}^{s - 1} (1 + k_i)\big] O\{k_t A(t)\}.
\end{align*}

On the other hand, similar to (\ref{equ1}) and (\ref{align1}) in \textbf{Step 2}, we have uniformly over $i \in J_{s}$,
\[\sum_{j \in J_{s + 1} \cup \cdots \cup J_{m + 1}} p_{ji}^2 \leq O\{A(s)\}.\]
Combining the above two bounds gives the convergence rate of $\sum_{j \in J_{s}} p_{ji}^2$ uniformly over $i \in J_{s}$ as
\begin{align*}
\sum_{j \in J_{s}} p_{ji}^2 \geq 1 - \sum_{t = 1}^{s - 1} \big[\prod_{i = t + 1}^{s - 1} (1 + k_i)\big] O\{k_t A(t)\} - O\{A(s)\}.
\end{align*}
Together with the induction assumption that our claim in (\ref{claim}) holds uniformly over $i \in J_l$, $1 \leq l \leq s - 1$, we know that the claim also holds uniformly over $i \in J_l$, $1 \leq l \leq s$. Therefore, by induction, the results in part (a) of Theorem \ref{Thmpred} hold uniformly over $i \in J_l$, $1 \leq l \leq m$.

\bigskip

\noindent \textbf{Proof of part (b).} In this part, we will show that when each group of spiked eigenvalues has size one (that is, $k_l = 1$ for any $l$, $1 \leq l \leq m$), the convergence rates of the angles between the sample score vectors $\bW \widehat{\bu}_i$ and the population score vectors $\bW \bu_i$, $1 \leq i \leq K$, are at least as fast as those of the angles between the corresponding sample and population eigenvectors established in part (a) of Theorem \ref{Thmpred}. The key idea is to conduct delicate analysis on the $\cos(\cdot)$ function of the angles between the sample score vectors and population score vectors, where some results about the sample eigenvalues derived in the proof of Lemma \ref{Thm1} will be used.

When each group has size one, we have $K = m$ and the convergence rates of $\widehat{\bu}_i$ ($i \in J_l$) to the space $span\{\bu_j: j \in J_l\}$ become the convergence rates of $\widehat{\bu}_i$ to $\bu_i$, $1 \leq i \leq K$. Denote by $\theta_{ii} = Angle(\widehat{\bu}_i, \bu_i)$ and $\omega_{ii} = Angle(\bW \widehat{\bu}_i, \bW \bu_i)$. Then the results in part (a) give that uniformly over $1 \leq i \leq K$,
\begin{align}\label{oii}
\cos^2(\theta_{ii}) =  p_{ii}^2 \geq 1 - \sum_{t = 1}^{i - 1} 2^{i - t - 1} O\{A(t)\} - O\{A(i)\}.
\end{align}
Since $\bS = n^{-1}\bW\t \bW = \widehat{\bU} \widehat{\bLambda} \widehat{\bU}\t$, $\widehat{\bu}_i$ would be the eigenvector of $\bW\t \bW$ corresponding to the eigenvalue $n \widehat{\lambda}_i$ with $L_2$-norm $1$. It follows that
\begin{equation*}\label{omega12}
\cos(\omega_{ii}) = \frac{(\bW \bu_i)\t \bW \widehat{\bu}_i}{\|\bW \bu_i\|_2 \|\bW \widehat{\bu}_i\|_2} = \frac{n \widehat{\lambda}_i \bu_i\t \widehat{\bu}_i}{\sqrt{n \widehat{\lambda}_i} \|\bW \bu_i\|_2} = \frac{\sqrt{n \widehat{\lambda}_i} \cos(\theta_{ii})}{\|\bW \bu_i\|_2}.
\end{equation*}
Squaring both sides above gives
\begin{align}\label{wii}
\cos^2(\omega_{ii}) = \frac{n \widehat{\lambda}_i \cos^2(\theta_{ii})}{\|\bW \bu_i\|_2^2}.
\end{align}
Therefore, it suffices to show $\|\bW \bu_i\|_2^2 \leq n \widehat{\lambda}_i$.

For the term $\|\bW \bu_i\|_2^2$, it follows from $\bW\t \bW = n \widehat{\bU} \widehat{\bLambda} \widehat{\bU}\t$ that
\begin{align*}
\|\bW \bu_i\|_2^2 = \bu_i\t \bW\t \bW \bu_i = n \bu_i\t \widehat{\bU} \widehat{\bLambda} \widehat{\bU}\t \bu_i = n \sum_{j = 1}^q \widehat{\lambda}_j (\bu_i\t \widehat{\bu}_j)^2 = n \sum_{j = 1}^q \widehat{\lambda}_j p_{ij}^2.
\end{align*}
By further making use of equality (\ref{pij}), we have
\begin{align*}
\|\bW \bu_i\|_2^2 = n\sum_{j = 1}^q \widehat{\lambda}_j p_{ij}^2 = \lambda_i \bz_i\t \bz_i,
\end{align*}
where $\bz_i$ is the $i$th column vector of $\bZ\t$. On the other hand, inequality (\ref{eig}) in the proof of Lemma \ref{Thm1} gives a lower bound for the sample eigenvalues $\widehat{\lambda}_i$, $1 \leq i \leq K$. Under the current setting that each group has size one, it gives
\[\widehat{\lambda}_i \geq \varphi_{1}(n^{-1} \lambda_i \bz_i \bz_i\t) = \varphi_{1}(n^{-1} \lambda_i \bz_i\t \bz_i) = n^{-1} \lambda_i \bz_i\t \bz_i,\]
where $\varphi_1 (\cdot)$ denotes the largest eigenvalue of a given matrix. It follows that
\[n \widehat{\lambda}_i \geq \lambda_i \bz_i\t \bz_i = \|\bW \bu_i\|_2^2.\]

Therefore, in view of (\ref{wii}), we get
\[\cos^2(\omega_{ii}) \geq \cos^2(\theta_{ii}),\]
which means that the convergence rate of the sample score vector is at least as good as that of the corresponding sample eigenvector. Then it follows from (\ref{oii}) that uniformly over $1 \leq i \leq K$,
\begin{align*}
\cos^2(\omega_{ii}) \geq 1 - \sum_{t = 1}^{i - 1} 2^{i - t - 1} O\{A(t)\} - O\{A(i)\},
\end{align*}
which completes the proof of part (b) of Theorem \ref{Thmpred}.

\subsection{Proof of Proposition \ref{L4}}

\smallskip

By Condition \ref{robXZ}, the inequality $\|n^{-1/2}(\bX, \bF) \bdelta\|_2 \geq c \|\bdelta\|_2$ holds for any $\bdelta$ satisfying $\|\bdelta\|_0 < M$ with significant probability $1 - \theta_{n,p}$. We now derive a similar result for $(\bX, \widehat{\bF})$ by analyzing the estimation errors of confounding factors $\bF$.

By the estimation error bound in Condition \ref{cond3}, we have for any $1 \leq j \leq K$,
\begin{align*}
\|\bff_j - \widehat{\bff}_j\|_2^2 & \leq \|\bff_j\|_2^2 + \|\widehat{\bff}_j\|_2^2 - 2 \ \bff_j' \widehat{\bff}_j = n + n - 2 \|\bff_j\|_2 \|\widehat{\bff}_j\|_2 \cos (\omega_{jj}) \\
& = 2n - 2n \cos (\omega_{jj}) = 2n \{1 - \cos (\omega_{jj})\} \leq \frac{c_2^2 \log n}{4 K^2 T^2}.
\end{align*}
Since the above bound does not vary with the index $j$, it gives the uniform confounding factor estimation error bound
\begin{equation}\label{pree}
\max_{1 \leq j \leq K}\|\bff_j - \widehat{\bff}_j\|_2 \leq \frac{c_2}{2K T}\sqrt{\log n}.
\end{equation}

Now we proceed to prove the inequality for $(\bX, \widehat{\bF})$. First of all, it follows from Condition \ref{robXZ} and the triangular inequality that
\begin{align*}
&\|n^{-1/2}(\bX, \widehat{\bF}) \bdelta\|_2 \geq \|n^{-1/2}(\bX, \bF) \bdelta\|_2 - \|n^{-1/2}(\bX, \bF) \bdelta - n^{-1/2}(\bX, \widehat{\bF}) \bdelta\|_2\\
&\geq  c \|\bdelta\|_2 - n^{-1/2}\|(\bF - \widehat{\bF}) \bdelta_1\|_2 \geq c \|\bdelta\|_2 - n^{-1/2} \max_{1 \leq j \leq K} \|(\bff_j - \widehat{\bff}_j)\|_2 \|\bdelta_1\|_1,
\end{align*}
where $\bdelta_1$ is a subvector of $\bdelta$ consisting of the last $K$ components. Note that $\|\bdelta_1\|_1 \leq \sqrt{K} \|\bdelta_1\|_2 \leq \sqrt{K} \|\bdelta\|_2$. Further applying inequality (\ref{pree}) yields
\[\|n^{-1/2}(\bX, \widehat{\bF}) \bdelta\|_2 \geq c \|\bdelta\|_2 - n^{-1/2} \cdot \frac{c_2}{2K T}\sqrt{\log n} \cdot \sqrt{K}\|\bdelta\|_2 \geq c_1 \|\bdelta\|_2,\]
where $c_1$ is some positive constant no larger than $c - \frac{c_2}{2 T} \sqrt{\frac{\log n}{n K}}$. It is clear that $c_1$ is smaller than but close to $c$ when $n$ is relatively large. In view of the tail probabilities in Conditions \ref{cond3} and \ref{robXZ}, the above inequality holds with probability at least $1 - \theta_1 - \theta_2$. Thus, we finish the proof of Proposition \ref{L4}.


\subsection{Proof of Theorem \ref{Thm4}}

\smallskip

With Proposition \ref{L4}, we will apply a similar idea as in \cite{Zheng2014} to prove the global properties. The proof consists of two parts. The first part shows the model selection consistency property with the range of $\lambda$ given in Theorem \ref{Thm4}. Based on the first part, several oracle inequalities will then be induced. We will first prove the properties when the columns of design matrix $\bX$ have a common scale of $L_2$-norm $n^{1/2}$ as a benchmark, meaning that $\bbeta_{\ast} = \bbeta$ and $L = 1$, and then illustrate the results in general cases.

\medskip

\noindent \textbf{Part 1: Model selection consistency.} This part contains two steps. In the first step, it will be shown that when $c_1^{-1}c_2\sqrt{(2s + 1)(\log p)/n} < \lambda < b_0$, the number of nonzero elements in $(\hbbeta\t, \hbgamma\t)\t$ is no larger than $s$ conditioning on the event $\widetilde{\mathcal{E}}$ defined in Lemma \ref{L3}. We prove this by using the global optimality of $(\hbbeta\t, \hbgamma\t)\t$.

By the hard-thresholding property \citep[Lemma 1]{Zheng2014} and $\lambda < b_0$, any nonzero component of the true regression coefficient vector $(\bbeta_0\t, \bgamma_0\t)\t$ or of the global minimizer $(\hbbeta\t, \hbgamma\t)\t$ is greater than $\lambda$, which ensures that $\|p_\lambda\{(\hbbeta\t, \hbgamma\t)\t\}\|_1 =  \lambda^2 \|(\hbbeta\t, \hbgamma\t)\t\|_0/2$ and $\|p_\lambda\{(\bbeta_0\t, \bgamma_0\t)\t\}\|_1= s \lambda^2/2$. Thus,
\[\Big\|p_\lambda\left\{(\hbbeta\t, \hbgamma\t)\t\right\}\Big\|_1 - \Big\|p_\lambda\left\{(\bbeta_0\t, \bgamma_0\t)\t\right\}\Big\|_1 = \left\{\|(\hbbeta\t, \hbgamma\t)\t\|_0 - s\right\}\lambda^2/2. \]
Denote by $\bdelta = (\hbbeta\t, \hbgamma\t)\t - (\bbeta_0\t, \bgamma_0\t)\t$.
Direct calculation yields
\begin{align}\label{simp}
Q\left\{(\hbbeta\t, \hbgamma\t)\t \right\} - Q\left\{(\bbeta_0\t, \bgamma_0\t)\t \right\} &=  2^{-1} \big\| n^{-\frac{1}{2}}(\bX, \widehat{\bF}) \bdelta \big\|_2^2 - n^{-1} \widetilde{\bveps}\t (\bX, \widehat{\bF}) \bdelta \nonumber\\
&+ \left\{\|(\hbbeta\t, \hbgamma\t)\t\|_0 - s \right\}\lambda^2/2,
\end{align}
where $\widetilde{\bveps} = (\bF - \widehat{\bF}) \bgamma + \bveps$, the sum of the random error vector $\bveps$ and estimation errors $(\bF - \widehat{\bF}) \bgamma$.

On the other hand, conditional on event $\widetilde{\mathcal{E}}$, we have
\begin{align}\label{E}
|n^{-1} \widetilde{\bveps}\t (\bX, \widehat{\bF}) \bdelta| \ &\leq \ \|n^{-1} \widetilde{\bveps}\t (\bX, \widehat{\bF})\|_{\infty}\|\bdelta\|_1 \  \\
&\leq \ c_2 \sqrt{(\log p)/n} \|\bdelta\|_1 \ \leq \ c_2 \sqrt{(\log p)/n}\|\bdelta\|_0^{\frac{1}{2}} \|\bdelta\|_2. \notag
\end{align}
In addition, by Condition \ref{cond2} and the definition of $\mathbb{S}_{M/2}$, we obtain $\|\bdelta\|_0 \leq \|(\bbeta_0\t, \bgamma_0\t)\t\|_0 + \|(\hbbeta\t, \hbgamma\t)\t\|_0 < M$, where $M$ is the robust spark of $(\bX, \widehat{\bF})$ with bound $c_1$ by Proposition \ref{L4}. Thus, we have
\begin{equation} \label{eigen}
\|n^{-\frac{1}{2}}(\bX, \widehat{\bF}) \bdelta\|_2 \geq c_1 \|\bdelta\|_2.
\end{equation}
Plugging inequalities (\ref{E}) and (\ref{eigen}) into (\ref{simp}) gives that
\begin{align} \label{Qbeta2}
Q \left\{ (\hbbeta\t, \hbgamma\t)\t \right\} - Q \left\{ (\bbeta_0\t, \bgamma_0\t)\t \right\} &\geq 2^{-1}c_1^2\|\bdelta\|_2^2 - \ c_2 \sqrt{(\log p)/n} \|\bdelta\|_0^{\frac{1}{2}} \|\bdelta\|_2 \nonumber\\
&+ \left\{ \|(\hbbeta\t, \hbgamma\t)\t\|_0 - s \right\} \lambda^2/2.
\end{align}

Thus, the global optimality of $(\hbbeta\t, \hbgamma\t)\t$ ensures that
\[2^{-1} c_1^2 \|\bdelta\|_2^2 - c_2 \sqrt{\frac{\log p}{n}} \|\bdelta\|_0^{\frac{1}{2}} \|\bdelta\|_2 + \left\{\|(\hbbeta\t, \hbgamma\t)\t\|_0 - s\right\} \lambda^2/2 \leq 0.\]
After completing the squares in the above inequality, we get
\begin{equation*}
\Big[c_1 \|\bdelta\|_2 - \frac{c_2}{c_1} \sqrt{\frac{\log p}{n}} \|\bdelta\|_0^{\frac{1}{2}}\Big]^2 - \left(\frac{c_2}{c_1}\right)^2 \frac{\log p}{n} \|\bdelta\|_0 + \left\{\|(\hbbeta\t, \hbgamma\t)\t\|_0 - s\right\}\lambda^2 \leq 0.
\end{equation*}
Since $ \Big[c_1 \|\bdelta\|_2 - \frac{c_2}{c_1} \sqrt{\frac{\log p}{n}} \|\bdelta\|_0^{\frac{1}{2}}\Big]^2\geq 0$, it gives
\begin{equation}\label{016}
\left\{\|(\hbbeta\t, \hbgamma\t)\t\|_0 - s\right\}\lambda^2 \leq \left(\frac{c_2}{c_1}\right)^2 \frac{\log p}{n} \|\bdelta\|_0.
\end{equation}

We continue to bound the value of $\|(\hbbeta\t, \hbgamma\t)\t\|_0$ by the above inequality. Let $k = \|(\hbbeta\t, \hbgamma\t)\t\|_0$. Then $\|\bdelta\|_0 = \|(\hbbeta\t, \hbgamma\t)\t - (\bbeta_0\t, \bgamma_0\t)\t\|_0 \leq k + s$. Thus, it follows from (\ref{016}) that
\[(k - s)\lambda^2 \leq \left(\frac{c_2}{c_1}\right)^2 \frac{\log p}{n} (k + s).\]
Organizing it in terms of $k$ and $s$, we get
\begin{equation} \label{equik}
k\left(\lambda^2 - \left(\frac{c_2}{c_1}\right)^2 \frac{\log p}{n}\right) \leq s\left(\lambda^2 + \left(\frac{c_2}{c_1}\right)^2 \frac{\log p}{n}\right).
\end{equation}

Since $\lambda > c_1^{-1}c_2\sqrt{(2s + 1)\log p/n}$, we have $\lambda^2 - (c_1^{-1}c_2)^2 (2 s + 1) \frac{\log p}{n} > 0$ and $\lambda^2 c_1^2 n - c_2^2 \log p > 2 c_2^2 s\log p$. Thus we have $ \frac{2c_2^2 \log p}{\lambda^2 c_1^2 n - c_2^2 \log p}<1/s$. Then it follows from inequality (\ref{equik}) that
\[k \leq s \frac{(\lambda^2 + (\frac{c_2}{c_1})^2 \frac{\log p}{n})}{(\lambda^2 - (\frac{c_2}{c_1})^2 \frac{\log p}{n})} = s \left(1 + \frac{2c_2^2 \log p}{\lambda^2 c_1^2 n - c_2^2 \log p}\right) < s + 1.\]
Therefore, the number of nonzero elements in $(\hbbeta\t, \hbgamma\t)\t$ satisfies
\[\big\|(\hbbeta\t, \hbgamma\t)\t \big\|_0 \leq s.\]

The second step is based on the first step, where we will use proof by contradiction to show that $\supp((\bbeta_0\t, \bgamma_0\t)\t) \subset \supp((\hbbeta\t, \hbgamma\t)\t)$ with the additional assumption $\lambda < b_0 c_1/\sqrt{2}$ in the theorem. Suppose that $\supp((\bbeta_0\t, \bgamma_0\t)\t) \not\subset \supp (\hbbeta\t, \hbgamma\t)\t$, and we denote the number of missed true coefficients as
\[
k = \left|\supp\big\{(\bbeta_0\t, \bgamma_0\t)\t \big\} \backslash \supp \big\{(\hbbeta\t, \hbgamma\t) \t\big\}\right| \geq 1.
\]
Then we have $\|(\hbbeta\t, \hbgamma\t)\t\|_0 \geq s - k$ and $\|\bdelta\|_0 \leq \|(\hbbeta\t, \hbgamma\t)\t\|_0 + \|(\bbeta_0\t, \bgamma_0\t)\t\|_0 \leq 2s$ by the first step. Combining these two results with  inequality (\ref{Qbeta2}) yields
\begin{align}\label{Q1}
Q \left\{ (\hbbeta\t, \hbgamma\t)\t \right\} - Q \left\{ (\bbeta_0\t, \bgamma_0\t)\t \right\} \geq \left(2^{-1} c_1^2 \|\bdelta\|_2 - c_2\sqrt{\frac{2s\log p}{n}}\right)\|\bdelta\|_2 - k \lambda^2/2.
\end{align}

Note that for each $j \in \supp((\bbeta_0\t, \bgamma_0\t)\t) \setminus \supp((\hbbeta\t, \hbgamma\t)\t)$, we have $|\delta_j|\geq b_0$ with $b_0$ the lowest signal strength defined in Condition \ref{cond2}. Thus, $\|\bdelta\|_2 \geq \sqrt{k}b_0$, which together with Condition \ref{cond2} entails
\[4^{-1} c_1^2 \|\bdelta\|_2 \geq 4^{-1} c_1^2 \sqrt{k}b_0 \geq 4^{-1} c_1^2 b_0 > c_2\sqrt{(2s\log p)/n}.\]
Thus, it follows from (\ref{Q1}) that
\begin{align*}
Q \left\{ (\hbbeta\t, \hbgamma\t)\t \right\} - Q \left\{ (\bbeta_0\t, \bgamma_0\t)\t \right\} \geq 4^{-1} c_1^2 \|\bdelta\|_2^2 - k \lambda^2/2 \geq  4^{-1} c_1^2 k b_0^2 - k \lambda^2/2 > 0,
\end{align*}
where the last step is because of the additional assumption $\lambda < b_0 c_1/\sqrt{2}$. The above inequality contradicts with the global optimality of $(\hbbeta\t, \hbgamma\t)\t$. Thus, we have $\supp((\bbeta_0\t, \bgamma_0\t)\t) \subset \supp((\hbbeta\t, \hbgamma\t)\t)$. Combining this with $\|(\hbbeta\t, \hbgamma\t)\t\|_0 \leq s$ from the first step, we know that $\supp \{(\hbbeta\t, \hbgamma\t)\t\} = \supp\{(\bbeta_0\t, \bgamma_0\t)\t\}$.

\medskip

\noindent \textbf{Part 2: Prediction and estimation losses}. In this part, we will bound the prediction and estimation losses. The idea is to get the $L_2$-estimation loss bound by the global optimality of $(\hbbeta\t, \hbgamma\t)\t$, conditional on the event $\widetilde{\mathcal{E}} \cap \widetilde{\mathcal{E}}_0$ defined in Lemma \ref{L3}. Then by similar techniques as in the first part, we would derive bounds for the prediction and estimation losses.

Recall that $\bX_0$, $\widehat{\bF}_0$ are the submatrices of $\bX$ and $\widehat{\bF}$ consisting of columns in $\supp(\bbeta_0)$ and $\supp(\bgamma_0)$, respectively. Conditioning on $\widetilde{\mathcal{E}} \cap \widetilde{\mathcal{E}}_0$, we have $\|\bdelta\|_0 \leq s$ by the model selection consistency established before. Thus, applying the Cauchy-Schwarz inequality and definition of $\widetilde{\mathcal{E}}_0$ gives
\begin{align} \label{018}
|n^{-1}\widetilde{\bveps}\t (\bX_0, \widehat{\bF}_0) \bdelta| \ &\leq \ \|n^{-1}\widetilde{\bveps}\t (\bX_0, \widehat{\bF}_0)\|_{\infty}\|\bdelta\|_1 \  \\
&\leq \ c_2 \sqrt{\frac{\log n}{n}} \|\bdelta\|_1 \ \leq \ c_2 \sqrt{\frac{s \log n}{n}} \|\bdelta\|_2. \notag
\end{align}
In views of (\ref{simp}) and (\ref{eigen}), it follows from inequality (\ref{018}) and the model selection consistency property $\|(\hbbeta\t, \hbgamma\t)\t\|_0 = s$ that
\begin{align*}
&Q \left\{ (\hbbeta\t, \hbgamma\t)\t \right\} - Q \left\{ (\bbeta_0\t, \bgamma_0\t)\t \right\}\\
= & \ 2^{-1} \|n^{-1}(\bX, \widehat{\bF})\bdelta\|_2^2 - n^{-1} \widetilde{\bveps}\t (\bX, \widehat{\bF}) \bdelta + \big\{\|(\hbbeta\t, \hbgamma\t)\t\|_0 - s\big\}\lambda^2/2\\
\geq & \ 2^{-1}c_1^2 \|\bdelta\|_2^2 - n^{-1} \widetilde{\bveps}\t (\bX_0, \widehat{\bF}_0) \bdelta \geq \Big(2^{-1} c_1^2 \|\bdelta\|_2 - c_2\sqrt{\frac{s\log n}{n}}\Big)\|\bdelta\|_2.
\end{align*}

Since $(\hbbeta\t, \hbgamma\t)\t$ is the global optimizer of $Q$, we have
\[2^{-1} c_1^2 \|\bdelta\|_2 - c_2\sqrt{\frac{s\log n}{n}} \leq 0,\]
which gives the $L_2$ and $L_\infty$ estimation loss bounds as
\begin{align*}
&\|(\hbbeta\t, \hbgamma\t)\t - (\bbeta_0\t, \bgamma_0\t)\t\|_2 = \|\bdelta\|_2 \leq \ 2c_1^{-2}c_2 \sqrt{(s\log n)/n}, \\
&\|(\hbbeta\t, \hbgamma\t)\t - (\bbeta_0\t, \bgamma_0\t)\t\|_{\infty} \leq \|(\hbbeta\t, \hbgamma\t)\t - (\bbeta_0\t, \bgamma_0\t)\t\|_2 \leq 2c_1^{-2}c_2 \sqrt{(s\log n)/n}.
\end{align*}
For $L_q$-estimation losses with $1 \leq q < 2$, applying H\"{o}lder's inequality gives
\begin{align*}
\nonumber \|(\hbbeta\t, \hbgamma\t)\t - &(\bbeta_0\t, \bgamma_0\t)\t\|_q = (\sum_{j} |\delta_j|^q)^{1/q} \leq (\sum_{j}|\delta_j|^2)^{\frac{1}{2}}(\sum_{\delta_j \neq 0} 1^{\frac{2}{2 - q}})^{\frac{1}{q}-\frac{1}{2}} \\
&= \|\bdelta\|_2 \|\bdelta\|_0^{\frac{1}{q}-\frac{1}{2}} \leq  2c_1^{-2}c_2 s^{\frac{1}{q}} \sqrt{(\log n)/n}.
\end{align*}

Next we prove the bound for oracle prediction loss. Since $(\hbbeta\t, \hbgamma\t)\t$ is the global minimizer, it follows from (\ref{simp}) and the model selection consistency property that
\begin{align*}
& n^{-1/2} \|(\bX, \widehat{\bF}) \{(\hbbeta\t, \hbgamma\t)\t - (\bbeta_0\t, \bgamma_0\t)\t\}\|_2 \\
\leq &\left\{2n^{-1} \widetilde{\bveps}\t (\bX, \widehat{\bF}) \bdelta \right\}^{1/2} \leq \left\{2\|n^{-1}(\bX_0, \widehat{\bF}_0)\t \tilde{\bveps}\|_\infty \|\bdelta\|_1\right\}^{1/2} \leq 2 c_2c_1^{-1} \sqrt{s (\log n)/n},
\end{align*}
where the last step is because of the $L_1$ estimation loss bound proved before. Then for the oracle prediction loss, together with (\ref{prederr}) in the proof of Lemma \ref{L3}, it follows that
\begin{align*}
&n^{-1/2} \|(\bX, \widehat{\bF}) (\hbbeta\t, \hbgamma\t)\t - (\bX, \bF) (\bbeta_0\t, \bgamma_0\t)\t\|_2 \\
\leq & \ 2 c_2c_1^{-1} \sqrt{s (\log n)/n} + n^{-1/2}\|(\bF - \widehat{\bF})\bgamma_0\|_2 \leq (2c_2c_1^{-1}\sqrt{s} + c_2/2) \sqrt{(\log n)/n}.
\end{align*}

\smallskip

Last we will derive our results for general cases when the $L_2$-norms of columns of $\bX$ are not of the common scale $n^{1/2}$. Note that the penalized least squares in (\ref{e001}) can be rewritten as
\begin{equation*}
Q\big\{(\bbeta\t, \bgamma\t)\t \big\} = (2n)^{-1} \|\by - \widetilde{\bX} \bbeta_{\ast} - \widehat{\bF} \bgamma\|_2^2 + \|p_\lambda\big\{(\bbeta_{\ast}\t, \bgamma\t)\t\big\}\|_1,
\end{equation*}
where $\widetilde{\bX}$ is the matrix with the $L_2$-norm of each column rescaled to $n^{1/2}$ and
$$\bbeta_{\ast}= n^{-1/2}(\beta_1 \|\bx_1\|_2, \dots, \beta_p \|\bx_p\|_2)\t$$
is the corresponding coefficient vector defined in (\ref{e001}). By Conditions \ref{condx} and \ref{cond2}, the same argument applies to derive the model selection consistency property and the bounds on oracle prediction and estimation losses for $(\hbbeta_{\ast}\t, \hbgamma\t)\t$ since the relationship between $\lambda$ and signal strength keeps the same even if $L \neq 1$. Based on Condition \ref{condx}, it is clear that the model selection consistency of $\widehat{\bbeta}_{\ast}$ implies that of $\widehat{\bbeta}$. And the bound on prediction loss does not change since $\widetilde{\bX} \hbbeta_{\ast} = \bX \hbbeta$. As for the bounds of estimation losses on $\widehat{\bbeta}$, they can be deduced as
\begin{align*}
&\|\hbbeta - \bbeta_0\|_2 \leq \ 2c_1^{-2}c_2L \sqrt{(s\log n)/n}, \ \ \|\hbbeta - \bbeta_0\|_q \leq 2c_1^{-2}c_2L s^{\frac{1}{q}} \sqrt{(\log n)/n},\\
&\|\hbbeta - \bbeta_0\|_{\infty} \leq \|\hbbeta - \bbeta_0\|_2 \leq 2L^{-1}c_1^{-2}c_2 \sqrt{(s\log n)/n}.
\end{align*}

The tail probability for these results to hold is at most the sum of the tail probabilities in Conditions \ref{cond3}-\ref{condx} and Lemma \ref{L3}. Thus, we know that these properties hold  simultaneously with probability at least
\begin{align*}
1 - \frac{4\sqrt{2}\sigma}{c_2 \sqrt{\pi \log p}} p^{1 - \frac{c_2^2}{8 \sigma^2}} + \frac{2\sqrt{2}\sigma s}{c_2 \sqrt{\pi \log n}} n^{-\frac{c_2^2}{8 \sigma^2}} - \theta_1 - \theta_2 - \theta_3.
\end{align*}
It concludes the proof of Theorem \ref{Thm4}.

\section{Additional technical details} \label{SecL}

The following lemma is needed in proving Lemma \ref{Thm1}.
\begin{lemma} [Weyl's inequality \citep{Horn1990}] \label{lem1}
If $\bA$ and $\bB$ are $m \times m$ real symmetric matrices, then for all $k = 1, \dots, m$,
\begin{equation*}
\left.\begin{aligned}
\varphi_k(\bA) &+ \varphi_m(\bB)\\
\varphi_{k + 1}(\bA) &+ \varphi_{m - 1}(\bB)\\
&\ \ \vdots\\
\varphi_m(\bA) &+ \varphi_k(\bB)
\end{aligned} \right\} \leq \varphi_k(\bA + \bB) \leq
\left\{\begin{aligned}
\varphi_k(\bA) &+ \varphi_1(\bB)\\
\varphi_{k - 1}(\bA) &+ \varphi_2(\bB)\\
&\ \ \vdots\\
\varphi_1(\bA) &+ \varphi_k(\bB)
\end{aligned} \right.,
\end{equation*}
where $\varphi_i(\cdot)$ is the function that takes the $i$th largest eigenvalue of a given matrix.
\end{lemma}

\subsection{Proof of Lemma \ref{Thm1}} \label{A.1}

\smallskip

The main idea of proving Lemma \ref{Thm1} is to use induction to show that the sample eigenvalues divided by their corresponding orders of $q$ will be convergent in an event with asymptotic probability one. To ease readability, the proof is divided into three steps.

\medskip

\noindent \textbf{Step 1: Large probability event $\mathcal{E}$.} In this step, we will define an event $\mathcal{E}$ and show that its probability approaches one when $q$ increases to infinity. Our later discussion will be conditional on this event. Denote a series of events by $\mathcal{E}_{jt}$, $1 \leq j \leq q$, $1 \leq t \leq n$, such that
\begin{align*}
\mathcal{E}_{jt} = \{z_{jt}^2 \leq K^{-1} q^{\alpha}\},
\end{align*}
where $z_{jt}$ is the $(j,t)$th entry of $\bZ$. By Condition \ref{tailpb}, the events $\mathcal{E}_{jt}$ satisfy a uniform tail probability bound $P(\mathcal{E}_{jt}^c) = o(q^{-1} n^{-1})$. Let $\mathcal{E} = \cap_{t = 1}^n \cap_{j = 1}^q \mathcal{E}_{jt}$ be the intersection of all events in the series. Then the probability of event $\mathcal{E}$ converges to one since
\begin{align*}
P(\mathcal{E}^c) & = P(\cup_{t = 1}^n \cup_{j = 1}^q \mathcal{E}_{jt}^c) \leq \sum_{t = 1}^n \sum_{j = 1}^q P(\mathcal{E}_{jt}^c) = n q \cdot o(q^{-1} n^{-1}) \to 0, \ as \ q \to \infty.
\end{align*}
\medskip

\noindent \textbf{Step 2: Convergence of eigenvalues with indices in $J_1$.} This is the first part of induction. We will show that conditional on event $\mathcal{E}$, uniformly over $i \in J_1$, $q^{-\alpha_1} \widehat{\lambda}_i \to c_i$, as $q \to \infty$.

Denote by $\bC$ the $q \times q$ diagonal matrix with the first $K$ diagonal components equaling to $c_j$, $1 \leq j \leq K$, and the rest diagonal components $1$. We decompose $\bZ,\bC$ and $\bLambda$ into block matrices according to the index sets $J_1, J_2, \dots, J_{m + 1}$ such that
\begin{equation}\label{C}
\bZ = \begin{pmatrix} \bZ_1\\ \bZ_2\\ \vdots \\\bZ_{m + 1}\end{pmatrix},
\bC = \begin{pmatrix}
&\bC_1 &\bO &\cdots &\bO\\
&\bO &\bC_2 &\cdots &\bO\\
&\vdots &\vdots &\ddots &\vdots\\
&\bO &\bO &\cdots &\bC_{m + 1}
\end{pmatrix},
\bLambda = \begin{pmatrix}
&\bLambda_1 &\bO &\cdots &\bO\\
&\bO &\bLambda_2 &\cdots &\bO\\
&\vdots &\vdots &\ddots &\vdots\\
&\bO &\bO &\cdots &\bLambda_{m + 1}
\end{pmatrix}.
\end{equation}
Then for the dual matrix $S_D$, we have
\begin{equation} \label{sd}
\bS_D = n^{-1}\bZ\t \bLambda \bZ =n^{-1}\sum_{l = 1}^{m + 1} \bZ_l\t \bLambda_l \bZ_l.
\end{equation}
Divided by $q^{\alpha_1}$ on both sides of (\ref{sd}) gives
\begin{align}\label{SDE}
q^{-\alpha_1}\bS_D &= n^{-1}q^{-\alpha_1} \bZ_1\t \bLambda_1 \bZ_1 + n^{-1}q^{-\alpha_1} \sum_{l = 2}^{m} \bZ_l\t \bLambda_l \bZ_l + n^{-1}q^{-\alpha_1} \bZ_{m + 1}\t \bLambda_{m + 1} \bZ_{m + 1}.
\end{align}
We will show the sum of the last two terms above converges to the zero matrix in Frobenius norm, where the Frobenius norm is defined as $\|\bA\|_F = \{\tr(\bA \bA\t)\}^{1/2}$ for a given matrix $\bA$.

For any $l$, $1 \leq l \leq m$, let $\lambda_t^{(l)}$ and $c_t^{(l)}$ be the $t$th diagonal elements of $\bLambda_l$ and $\bC_l$, respectively. Conditional on event $\mathcal{E}$, for any $j$ and $k$, $1 \leq j,k \leq n$, the absolute value of the $(j,k)$th element in $\sum_{l = 2}^{m} \bZ_l\t \bLambda_l \bZ_l$ is
\begin{align*}
|\sum_{l = 2}^{m} \sum_{t = 1}^{k_l} \lambda_t^{(l)} {z_{tj}^{(l)}} z_{tk}^{(l)}| \leq K^{-1} q^{\alpha} \sum_{l = 2}^{m} \sum_{t = 1}^{k_l} \lambda_t^{(l)},
\end{align*}
where ${z_{tj}^{(l)}}$ and $z_{tk}^{(l)}$ are the $(t,j)$th and $(t,k)$th elements in $\bZ_l$, respectively. By Condition \ref{eigenvl}, uniformly over $1 \leq l \leq m$ and $1 \leq t \leq k_l$, $\lambda_t^{(l)} = O(q^{\alpha_l} c_t^{(l)})$. Then it follows that
\begin{align*}
\|&n^{-1}q^{-\alpha_1} \sum_{l = 2}^{m} \bZ_l\t \bLambda_l \bZ_l\|_F \leq n^{-1} q^{-\alpha_1} (n K^{-1} q^{\alpha} \sum_{l = 2}^{m} \sum_{t = 1}^{k_l} \lambda_t^{(l)})\\
= & \ O\{q^{-\alpha_1} K^{-1} q^{\alpha} \sum_{l = 2}^{m} \sum_{t = 1}^{k_l} q^{\alpha_l} c_t^{(l)}\} = O\{K^{-1} q^{\alpha} \sum_{l = 2}^{m} k_l C q^{\alpha_l}/q^{\alpha_1}\}.
\end{align*}
Similarly we would get
\[\|n^{-1}q^{-\alpha_1} \bZ_{m + 1}\t \bLambda_{m + 1} \bZ_{m + 1}\|_F \leq K^{-1} q^{\alpha} k_{m + 1} C/q^{\alpha_1}.\]
Together with $\alpha < \mbox{min}\{\Delta, \alpha_m - 1\}$ by Condition \ref{tailpb} and $k_{m + 1} < q$, we have
\begin{align}\label{Fconv0}
&\|n^{-1}q^{-\alpha_1} \sum_{l = 2}^{m} \bZ_l\t \bLambda_l \bZ_l + n^{-1}q^{-\alpha_1} \bZ_{m + 1}\t \bLambda_{m + 1} \bZ_{m + 1}\|_F \nonumber\\
\leq & \ O\{(\sum_{l = 2}^{m} k_l q^{\alpha_l} + k_{m + 1}) K^{-1} q^{\alpha} C/q^{\alpha_1}\} \to 0, \ as \ q \to \infty.
\end{align}

By a similar argument, under Condition \ref{eigenvl}, we have
\begin{align}\label{convZ1}
&\|n^{-1}q^{-\alpha_1} \bZ_1\t \bLambda_1 \bZ_1 - n^{-1}\bZ_1\t\bC_1 \bZ_1\|_F = \|n^{-1} \bZ_1\t (q^{-\alpha_1}\bLambda_1 - \bC_1) \bZ_1\|_F \nonumber\\
\leq \ &n^{-1} [n K^{-1} q^{\alpha} \sum_{t = 1}^{k_1}(q^{-\alpha_1} \lambda_t^{(1)} - c_t^{(1)})] \leq  k_1 K^{-1} q^{\alpha} \cdot O(q^{-\Delta}) \to 0, \ as \ q \to \infty.
\end{align}
In view of (\ref{SDE}), it is immediate that
\begin{align}\label{Fconv}
&\|q^{-\alpha_1}\bS_D - n^{-1}\bZ_1\t\bC_1 \bZ_1\|_F \leq \|n^{-1}q^{-\alpha_1} \bZ_1\t \bLambda_1 \bZ_1 - n^{-1}\bZ_1\t\bC_1 \bZ_1\|_F \nonumber\\
& + \|n^{-1}q^{-\alpha_1} \sum_{l = 2}^{m} \bZ_l\t \bLambda_l \bZ_l + n^{-1}q^{-\alpha_1} \bZ_{m + 1}\t \bLambda_{m + 1} \bZ_{m + 1}\|_F \to 0, \ as \ q \to \infty.
\end{align}
Further applying \cite[Corollary 6.3.8]{Horn1990} gives as $q \to \infty$,
\begin{align}\label{igenbd}
\max_{1 \leq i \leq n}|\varphi_i(q^{-\alpha_1}\bS_D) - \varphi_i(n^{-1}\bZ_1\t\bC_1 \bZ_1)| \leq \|q^{-\alpha_1}\bS_D - n^{-1}\bZ_1\t\bC_1 \bZ_1\|_F \to 0.
\end{align}

Note that $n^{-1}\bZ_1\t\bC_1 \bZ_1$ shares the same nonzero eigenvalues with its due matrix $n^{-1}\bC_1^{1/2}\bZ_1 \bZ_1\t \bC_1^{1/2}$ of dimensionality $k_1$. It follows from (\ref{igenbd}) that
\begin{align}\label{igenbd2}
\max_{i \in J_1}|\varphi_i(q^{-\alpha_1}\bS_D) - \varphi_i(n^{-1}\bC_1^{1/2}\bZ_1 \bZ_1\t \bC_1^{1/2})| \to 0.
\end{align}
Moreover, by part (b) of Condition \ref{tailpb}, we have
\begin{align}\label{igenbd3}
\max_{i \in J_1}|\varphi_i(n^{-1}\bC_1^{1/2}\bZ_1 \bZ_1\t \bC_1^{1/2}) - \varphi_i(\bC_1)| \leq \|\bC_1^{1/2} (n^{-1}\bZ_1 \bZ_1\t - \bI_{k_1}) \bC_1^{1/2}\|_F \to 0.
\end{align}
Therefore, (\ref{igenbd2}) and (\ref{igenbd3}) together yield that uniformly over $i \in J_1$,
\begin{align*}
q^{-\alpha_1} \widehat{\lambda}_i = \varphi_i(q^{-\alpha_1}\bS_D) \to \varphi_i(\bC_1) = c_i,
\end{align*}
as $q \to \infty$. It completes the proof of \textbf{Step 2}.

\medskip

\noindent \textbf{Step 3: Convergence of eigenvalues with indices in $J_2, \dots, J_m$.} As the second part of induction, for any fixed $t$, $2 \leq t \leq m$, we will show $q^{-\alpha_t} \widehat{\lambda}_i \to c_i$ for any $i \in J_t$, as $q \to \infty$. The basic idea in this step is to use Weyl's inequality (Lemma \ref{lem1}) to get both a lower bound and an upper bound of $q^{-\alpha_t} \widehat{\lambda}_i$, and show that they converge to the same limit.

We derive the upper bound first. Divided by $q^{\alpha_t}$ on both sides of (\ref{sd}) gives
\[q^{-\alpha_t}\bS_D = n^{-1}q^{-\alpha_t} \sum_{l = 1}^{t - 1} \bZ_l\t \bLambda_l \bZ_l + n^{-1}q^{-\alpha_t} \sum_{l = t}^{m + 1} \bZ_l\t \bLambda_l \bZ_l.\]
Applying Weyl's inequality, we get
\begin{align}\label{upbound}
\varphi_i(q^{-\alpha_t}\bS_D) &\leq \varphi_{1 + \sum_{l = 1}^{t - 1} k_l}( \sum_{l = 1}^{t - 1} \bZ_l\t \bLambda_l \bZ_l / nq^{\alpha_t}) + \varphi_{i - \sum_{l = 1}^{t - 1} k_l}(n^{-1}q^{-\alpha_t} \sum_{l = t}^{m + 1} \bZ_l\t \bLambda_l \bZ_l) \nonumber\\
&= \varphi_{i - \sum_{l = 1}^{t - 1} k_l}(n^{-1}q^{-\alpha_t} \sum_{l = t}^{m + 1} \bZ_l\t \bLambda_l \bZ_l),
\end{align}
where the first term is indeed zero since $n^{-1}q^{-\alpha_t} \sum_{l = 1}^{t - 1} \bZ_l\t \bLambda_l \bZ_l$ has a rank no more than $\sum_{l = 1}^{t - 1} k_l$. It gives an upper bound of $\varphi_i(q^{-\alpha_t}\bS_D)$. By the same argument as (\ref{Fconv}) in \textbf{Step 2}, under Conditions \ref{eigenvl} and \ref{tailpb}, we have
\[\|n^{-1}q^{-\alpha_t} \sum_{l = t}^{m + 1} \bZ_l\t \bLambda_l \bZ_l - n^{-1} \bZ_t\t \bC_t \bZ_t\|_F \to 0, \ as \ q \to \infty.\]
Similar to (\ref{igenbd}), it implies the upper bound of $\varphi_i(q^{-\alpha_t}\bS_D)$ in (\ref{upbound}) converges to the same limit as $\varphi_{i - \sum_{l = 1}^{t - 1} k_l}(n^{-1}\bZ_t\t\bC_t \bZ_t)$ uniformly over $i \in J_t$ as $q \to \infty$.

On the other hand, by Weyl's inequality, we also have
\begin{align*}
\varphi_i(q^{-\alpha_t}\bS_D) &\geq \varphi_i(n^{-1}q^{-\alpha_t} \sum_{l = 1}^t \bZ_l\t \bLambda_l \bZ_l) + \varphi_n(n^{-1}q^{-\alpha_t} \sum_{l = t + 1}^{m + 1} \bZ_l\t \bLambda_l \bZ_l)\\
& \geq \varphi_i(n^{-1}q^{-\alpha_t} \sum_{l = 1}^t \bZ_l\t \bLambda_l \bZ_l),
\end{align*}
where the second term vanishes since the eigenvalues of $\sum_{l = t + 1}^{m + 1} \bZ_l\t \bLambda_l \bZ_l$ are non-negative. In fact, $n^{-1}q^{-\alpha_t} \sum_{l = t + 1}^{m + 1} \bZ_l\t \bLambda_l \bZ_l$ would converge to a zero matrix in Frobenius norm under Conditions \ref{eigenvl} and \ref{tailpb}, similarly as in (\ref{Fconv0}). For the term $\varphi_i(n^{-1}q^{-\alpha_t} \sum_{l = 1}^t \bZ_l\t \bLambda_l \bZ_l)$, we use Weyl's inequality once more to get
\begin{align*}
&\varphi_{\sum_{l = 1}^t k_l}(n^{-1}q^{-\alpha_t} \sum_{l = 1}^{t - 1} \bZ_l\t \bLambda_l \bZ_l) \\
\leq \ &\varphi_i(n^{-1}q^{-\alpha_t} \sum_{l = 1}^t \bZ_l\t \bLambda_l \bZ_l) + \varphi_{1 - i + \sum_{l = 1}^t k_l}(-n^{-1}q^{-\alpha_t} \bZ_t\t \bLambda_t \bZ_t).
\end{align*}
Note that the term on the left hand side is indeed zero since the inside matrix has a rank no more than $\sum_{l = 1}^{t - 1} k_l$. It follows that
\begin{align*}
&\varphi_i  (n^{-1}q^{-\alpha_t} \sum_{l = 1}^t \bZ_t\t \bLambda_t \bZ_t)\geq -\varphi_{1 - i + \sum_{l = 1}^t k_l}(-n^{-1}q^{-\alpha_t} \bZ_t\t \bLambda_t \bZ_t) \\
= &\varphi_{k_t - (1 - i + \sum_{l = 1}^t k_l) + 1}(n^{-1}q^{-\alpha_t} \bZ_t\t \bLambda_t \bZ_t) = \varphi_{i - \sum_{l = 1}^{t - 1} k_l}(n^{-1}q^{-\alpha_t} \bZ_t\t \bLambda_t \bZ_t),
\end{align*}
where we make use of the fact that $\varphi_i(\bA) = - \varphi_{n - i + 1}(-\bA)$ for any $n \times n$ real symmetric matrix $\bA$, and any $1 \leq i \leq n$.

Therefore, we get a lower bound $\varphi_{i - \sum_{l = 1}^{t - 1} k_l}(n^{-1} q^{-\alpha_t} \bZ_t\t \bLambda_t \bZ_t)$ for $\varphi_i(q^{-\alpha_t} \bS_D)$. In terms of sample eigenvalues, the above argument shows that for any $\widehat{\lambda}_i$, $i \in J_t$, $1 \leq t \leq m$,
\begin{equation} \label{eig}
\widehat{\lambda}_i = \varphi_i(\bS_D) \geq \varphi_{i - \sum_{l = 1}^{t - 1} k_l}(n^{-1} \bZ_t\t \bLambda_t \bZ_t),
\end{equation}
which is useful in proving the convergence properties of the sample score vectors.

Now we show that the two bounds converge to the same limit. Similar to (\ref{convZ1}), as $q \to \infty$, we have
\[\|n^{-1}q^{-\alpha_t} \bZ_t\t \bLambda_t \bZ_t - n^{-1} \bZ_t\t \bC_t \bZ_t\|_F \to 0,\]
which gives
\[\max_{i \in J_t} \big|\varphi_{i - \sum_{l = 1}^{t - 1} k_l}(n^{-1}q^{-\alpha_t} \bZ_t\t \bLambda_t \bZ_t) - \varphi_{i - \sum_{l = 1}^{t - 1} k_l}(n^{-1} \bZ_t\t \bC_t \bZ_t)\big| \to 0.\]
It shows that the lower bound of $\varphi_i(q^{-\alpha_t} \bS_D)$ converges to the same limit as $\varphi_{i - \sum_{l = 1}^{t - 1} k_l}(n^{-1}\bZ_t\t\bC_t \bZ_t)$ uniformly over $i \in J_t$, so does the upper bound in (\ref{upbound}). It follows that $\varphi_i(q^{-\alpha_t}\bS_D)$ would also converge to the same limit as $\varphi_{i - \sum_{l = 1}^{t - 1} k_l}(n^{-1}\bZ_t\t\bC_t \bZ_t)$ uniformly over $i \in J_t$. That is, as $q \to \infty$,
\begin{align*}
\max_{i \in J_t}|\varphi_i(q^{-\alpha_t}\bS_D) - \varphi_{i - \sum_{l = 1}^{t - 1} k_l}(n^{-1}\bZ_t\t\bC_t \bZ_t)| \to 0.
\end{align*}
By a similar argument as in (\ref{igenbd2}) and (\ref{igenbd3}), we then have
\begin{align*}
\varphi_i(q^{-\alpha_t}\bS_D) \to \varphi_{i - \sum_{l = 1}^{t - 1} k_l}(\bC_t) = c_i,
\end{align*}
uniformly over $i \in J_t$, as $q \to \infty$. Along with the first step of induction in \textbf{Step 2}, we finish the proof of Lemma \ref{Thm1}.



\subsection{Proof of Lemma \ref{L3}}

\smallskip

To prove the probability bound in Lemma \ref{L3}, we will apply Bonferroni's inequality and Gaussian tail probability bound. Since $\widetilde{\bveps} = (\bF - \widehat{\bF}) \bgamma + \bveps$, some important bounds are needed before continuation. First, the inequality $\|\bgamma\|_1 \leq K T$ follows immediately from the fact $\|\bgamma\|_{\infty} \leq T$. Moreover, based on the estimation error bound of $\widehat{\bF}$ in Condition \ref{cond3}, we know that inequality (\ref{pree}) holds. These two inequalities yield
\begin{equation}\label{prederr}
\|(\bF - \widehat{\bF}) \bgamma\|_2 \leq \|\bgamma\|_1 \cdot \max_{1 \leq j \leq K}\|\bff_j - \widehat{\bff}_j\|_2 \leq K T \cdot \frac{c_2}{2K T}\sqrt{\log n} = \frac{c_2}{2} \sqrt{\log n},
\end{equation}
which gives
\begin{align}\label{XF}
n^{-1}|\bx_i\t (\bF - \widehat{\bF}) \bgamma| \leq n^{-1/2} \|(\bF - \widehat{\bF}) \bgamma\|_2 \leq \frac{c_2}{2} \sqrt{\frac{\log n}{n}}\leq \frac{c_2}{2} \sqrt{\frac{\log p}{n}}.
\end{align}
Similarly we have $n^{-1}|\bff_j\t (\bF - \widehat{\bF}) \bgamma| \leq 2^{-1}c_2 \sqrt{(\log n)/n}$.

Now we proceed to prove the probability bounds of the two events. Recall that both $\bff_j$ and $\widehat{\bff}_j$ have been rescaled to have $L_2$-norm $n^{1/2}$ and $\bveps \sim N(\textbf{0},\sigma^2 \bI_n)$ (Section \ref{Sec2}). Given $\bx_i$ and $\widehat{\bff}_j$, it follows that $n^{-1} \bx_i\t \bveps \sim N(0, \sigma^2 /n)$ and $n^{-1} \widehat{\bff}_j\t \bveps \sim N(0, \sigma^2/n)$ for any $i$ and $j$. By Bonferroni's inequality, the tail probability of $\widetilde{\mathcal{E}}$ satisfies 
\begin{align*}
P(\widetilde{\mathcal{E}}^c) \leq \sum^{p}_{i = 1} P \left(|n^{-1} \bx_i\t \widetilde{\bveps}| > c_2 \sqrt{(\log p)/n} \right) + \sum^{K}_{j = 1} P \left(|n^{-1} \widehat{\bff}_j\t \widetilde{\bveps}| > c_2 \sqrt{(\log p)/n} \right).
\end{align*}
By inequality (\ref{XF}) and Gaussian tail probability bound, for the first term on the right hand side above, we have
\begin{align*}
\sum^{p}_{i = 1} P \left( \frac{|\bx_i\t \widetilde{\bveps}|}{n} > c_2 \sqrt{\frac{\log p}{n}} \right) \leq \sum^{p}_{i = 1} P \left(\frac{|\bx_i\t \bveps|}{n} > c_2 \sqrt{\frac{\log p}{n}} - n^{-1} |\bx_i\t (\bF - \widehat{\bF}) \bgamma| \right)\\
\leq \sum^{p}_{i = 1} P \left(\frac{|\bx_i\t \bveps|}{n} > \frac{c_2}{2} \sqrt{\frac{\log p}{n}} \right)
\leq \sum^{p}_{j = 1} \frac{4 \sigma}{c_2 \sqrt{\log p}} \frac{1}{\sqrt{2 \pi}} e^{-\frac{c_2^2 \log p}{8 \sigma^2}} \leq \frac{2\sqrt{2}\sigma}{c_2 \sqrt{\pi \log p}} p^{1 - \frac{c_2^2}{8 \sigma^2}}.
\end{align*}
For the second term, similarly we have
\begin{align*}
\sum^{K}_{j = 1} P \Big (|n^{-1} \widehat{\bff}_j\t \widetilde{\bveps}| > c_2 \sqrt{(\log p)/n} \Big ) \leq \frac{2\sqrt{2}\sigma K}{c_2 \sqrt{\pi \log p}} p^{-\frac{c_2^2}{8 \sigma^2}}.
\end{align*}
As $K$ is no larger than $p$, the two bounds above give
\[P(\widetilde{\mathcal{E}}^c) \leq \frac{4\sqrt{2}\sigma}{c_2 \sqrt{\pi \log p}} p^{1 - \frac{c_2^2}{8 \sigma^2}}.\]

By a similar argument, the bound on $P(\widetilde{\mathcal{E}}_0^c)$ can be derived as
\[P(\widetilde{\mathcal{E}}_0^c) \leq  \frac{2\sqrt{2}\sigma s}{c_2 \sqrt{\pi \log n}} n^{-\frac{c_2^2}{8 \sigma^2}}.\]
Thus, for the intersection event $\widetilde{\mathcal{E}} \cap \widetilde{\mathcal{E}}_0$, we have
\begin{align*}
P\{(\widetilde{\mathcal{E}} \cap \widetilde{\mathcal{E}}_0)^c\} \leq P(\tilde{\mathcal{E}}^c) + P(\widetilde{\mathcal{E}}_0^c) \leq \frac{4\sqrt{2}\sigma}{c_2 \sqrt{\pi \log p}} p^{1 - \frac{c_2^2}{8 \sigma^2}} + \frac{2\sqrt{2}\sigma s}{c_2 \sqrt{\pi \log n}} n^{-\frac{c_2^2}{8 \sigma^2}},
\end{align*}
which converges to zero as $n\to \infty$ for $c_2 > 2\sqrt{2} \sigma$. It completes the proof of Lemma \ref{L3}.

\end{document}